\documentclass[aps,prd,onecolumn,amsmath,amssymb,nofootinbib,11pt]{revtex4-2}

\usepackage{bm}
\usepackage{graphicx}
\usepackage{hyperref}
 \hypersetup{
     colorlinks=true,
     linkcolor=black,
     filecolor=blue,
     citecolor=blue,
     urlcolor=cyan,
     pdfencoding=auto,
     pdftitle={Geodesics of the five-dimensional Kerr-(A)dS black hole in closed form},
     pdfauthor={\"Ozg\"ur Sar\i o\u{g}lu, Enes Y\i lmaz},
     }

\newcommand{\Xir}{\Xi_r}                  % Xi_r = 1 - lambda r^2
\newcommand{\Xii}{\Xi_i}                  % Xi_i = 1 + lambda a_i^2
\newcommand{\Xij}{\Xi_j}
\newcommand{\Xia}{\Xi_a}
\newcommand{\Xib}{\Xi_b}
\newcommand{\Ri}{R_i}                     % R_i  = r^2 + a_i^2
\newcommand{\Rj}{R_j}
\newcommand{\rhs}{\rho^{2}}               % rho^2
\newcommand{\chit}{\chi}                  % chi   = a^2 cos^2 th + b^2 sin^2 th
\newcommand{\Dth}{\Delta_\theta}          % Delta_theta
\newcommand{\Kkl}{K_{\rm KL}}             % the separation constant used in KL
\newcommand{\kap}{\kappa}                 % kappa     = 2MV/(rho^2 (V-2M))
\newcommand{\kah}{\widehat{\kappa}}       % hat kappa = rho^2 kappa = 2MV/(V-2M)
\newcommand{\Ecal}{{\mathcal E}}          % E - lambda a_i L_i
\newcommand{\eps}{\varepsilon}            % -1 timelike, 0 null
\newcommand{\Rcal}{{\mathcal R}}          % radial  potential of the first-order system
\newcommand{\Tcal}{{\mathcal T}}          % polar   potential of the first-order system
\newcommand{\Kcal}{{\mathcal K}}          % K / r_0^2, the rescaled separation constant
\newcommand{\ph}{\varphi}
\newcommand{\dd}{\mathrm{d}}
\newcommand{\half}{\tfrac{1}{2}}          % INLINE MATH ONLY -- in displayed
\allowdisplaybreaks[4]

\newcommand{\arxiv}[1]{%
  \mbox{[\href{https://arxiv.org/abs/#1}{\textcolor{magenta}{\ttfamily arXiv:#1}}]}}

\DeclareTextCommand{\textunderscore}{OT1}{\leavevmode{\ttfamily\char95}}

\begin{document}

%%  \texorpdfstring keeps the \\ out of the PDF-metadata form of the title,
%%  which hyperref builds from \title and cannot break
\title{Geodesics of the five-dimensional Kerr--(A)dS black hole%
       \texorpdfstring{\\}{ }in closed form}

\author{{\"O}zg{\"u}r Sar{\i}o\u{g}lu}
\email{sarioglu@metu.edu.tr}
\affiliation{Department of Physics, Faculty of Arts and Sciences,\\
              Middle East Technical University, 06800, Ankara, Turkey}

\author{Enes Y{\i}lmaz}
%%  the \_ here needs the OT1 redefinition in the preamble; see the note there
\email{enes.yilmaz\_02@metu.edu.tr}
\affiliation{Department of Physics, Faculty of Arts and Sciences,\\
              Middle East Technical University, 06800, Ankara, Turkey}

%%  A fixed date, not \today: arXiv rebuilds the PDF from source, and \today
%%  would then print the date of the rebuild rather than the date of writing.
%%  Update this line if the submission date changes.
\date{October 19, 2026}

\begin{abstract}
The Hamilton--Jacobi equation for geodesic motion in the five-dimensional Kerr--(A)dS black
hole separates in Boyer--Lindquist coordinates, but the resulting first-order system has so
far been integrated case by case, according to the real roots of the radial and the polar
polynomial. We give the general solution in closed form. In dimensionless variables both
polynomials are quartic, and a theorem of Biermann and Weierstrass then yields the radial
and the polar coordinate as one pair of manifestly real formulas, parametrized by an
arbitrary point of the orbit rather than by a turning point, and valid for timelike and
null geodesics alike whatever the root structure. The time coordinate and the two azimuthal
angles follow by quadratures that reduce to elliptic integrals of the third kind alone, and
the affine parameter by one more of the same type. Along the way we correct the inverse
metric and the radial potential as published.
\end{abstract}

\maketitle

%%%%%%%%%%%%%%%%%%%%%%%%%%%%%%%%%%%%%%%%%%%%%%%%%%%%%%%%%%%%%%%%%%%%%%%%%%%%%%
%%  I.  INTRODUCTION
%%%%%%%%%%%%%%%%%%%%%%%%%%%%%%%%%%%%%%%%%%%%%%%%%%%%%%%%%%%%%%%%%%%%%%%%%%%%%%
\section{Introduction}\label{sec:intro}

For obvious reasons the timelike and null geodesics of the Kerr spacetime matter a great
deal in astrophysics. What is comparatively recent is that they can be written down
analytically, in closed form, in terms of Weierstrass elliptic functions
\cite{Hackmann:thesis,Cieslik:2022uki,Cieslik:2023qdc}. The shape of that result is as
notable as the result itself. A single set of formulas covers every physical geodesic, the
expressions are explicitly real, and the initial data enter as free parameters, so that
nothing has to be split into cases according to the root structure of the underlying
polynomial. The instrument behind it is a little-known theorem of Biermann and Weierstrass
\cite{Biermann:1865,WhittakerWatson,Greenhill:1892}, which solves
$(\dd y/\dd z)^{2} = f(y)$ for a quartic $f$ about an arbitrary point of the solution
rather than about a zero of $f$.

Meanwhile it is well known \cite{Frolov:2017kze,Kunduri:2005fq} that the Hamilton--Jacobi
equation is separable for the five-dimensional Kerr--(A)dS metric in Boyer--Lindquist form,
which gives Liouville integrability and a first-order form of the geodesic equations. Our
initial motivation was to examine whether closed-form analytical solutions as in the Kerr
case could be constructed here as well, and to see what differences would emerge compared
to the four-dimensional case. After carrying the separation through and finding that it
yields quartic polynomials in the radial and in the polar sector alike, so that the
Biermann--Weierstrass theorem applies to both, we became aware of the work of Grunau,
Neumann and Reimers \cite{Grunau:2017uwv}, who had studied the same geodesics in detail for
a negative cosmological constant and integrated them case by case. We continued, and the
present paper is what came of it.

Four things are reported here. The first is a correction. The $g^{\ph_i\ph_j}$ block of the
inverse metric as published in \cite{Kunduri:2005fq} is wrong, and the error propagates
into the radial potential of the separated system and from there into the indicial
exponents of the radial Klein--Gordon equation at the horizon. With the corrected potential
those exponents return the superradiance combination $\omega - \Omega_i m_i$ automatically;
with the published one they do not. The second is a statement about degrees. In the
dimensionless variables of Sec.~\ref{sec:rescaled} both orbital polynomials are quartic,
with leading coefficients proportional to the product of the cosmological constant and the
causal normalization $\eps$. Because the natural radial variable in five dimensions is
$r^{2}$, the cosmological constant raises the radial degree only from three to four, where
in four-dimensional Kerr--(A)dS it raises it from four to six. Both sectors therefore stay
elliptic in five dimensions for timelike geodesics as well as null, while their
four-dimensional counterparts do not.

The third is the solution itself: a single pair of
formulas for the radial and the polar coordinate, manifestly real, valid whatever the root
structure of the two polynomials, and parametrized by an arbitrary point of the orbit
rather than by a turning point. That last freedom is not a convenience. For a positive
cosmological constant the radial quartic can be positive for every value of its argument,
and the orbit then has no turning point at which the classical substitution could be
started at all. The fourth is that the quadratures for the time and for the two azimuthal
angles reduce to elliptic integrals of the third kind alone, with residues in closed form.
The affine parameter, which the Mino parameter replaces along the way, comes back by one
further quadrature of the same type.
Those residues are proportional to the superradiance bracket $E - \Omega_i(x_h)\ell_i$ and
stand in the fixed ratio $\Omega_i(x_h)$ between an azimuthal angle and the time, which
accounts, in closed form, for the divergences that attend a horizon crossing in these
coordinates.

The paper is organized as follows. Section~\ref{sec:metric} records the metric in
Boyer--Lindquist form, inverts it and compares the inverse with the published one.
Section~\ref{sec:hj} separates the Hamilton--Jacobi equation, with the separation constant
fixed so that every expression stays regular as the cosmological constant goes to zero.
Section~\ref{sec:firstorder} turns the result into a first-order system, reparametrizes it
by the Mino parameter, passes to dimensionless variables and extracts the two quartics. It
closes with a detailed comparison with \cite{Grunau:2017uwv}. Section~\ref{sec:bw} is the
solution: the Biermann--Weierstrass theorem in Sec.~\ref{sec:bwthm}, the radial and polar
coordinates in Secs.~\ref{sec:radialsol} and \ref{sec:polarsol}, the time coordinate and
the azimuthal angles in Secs.~\ref{sec:timesol} and \ref{sec:azimuthalsol}, the affine
parameter in Sec.~\ref{sec:affine}, and the degenerate cases in
Sec.~\ref{sec:degenerate}. Section~\ref{sec:conclusions} collects the
conclusions and a few directions that seem worth pursuing. Appendix~\ref{app:kg} records
what the correction of Sec.~\ref{sec:compare} does to the Klein--Gordon equation.

%%%%%%%%%%%%%%%%%%%%%%%%%%%%%%%%%%%%%%%%%%%%%%%%%%%%%%%%%%%%%%%%%%%%%%%%%%%%%%
%%  II.  THE METRIC AND ITS INVERSE
%%%%%%%%%%%%%%%%%%%%%%%%%%%%%%%%%%%%%%%%%%%%%%%%%%%%%%%%%%%%%%%%%%%%%%%%%%%%%%
\section{The metric and its inverse}\label{sec:metric}

Before any of the dynamics we need the geometry in a usable form. This section records the
five-dimensional Kerr--(A)dS metric in Boyer--Lindquist coordinates, fixes the notation
used throughout and collects the structure functions that will recur. It then inverts the
metric. The inverse is where the separability of Sec.~\ref{sec:hj} can be read off, and it
is also where the published expressions stand in need of a correction.

%%-------------------------------------------------------------------- II.A --
\subsection{The Boyer--Lindquist form and its structure functions}\label{sec:metricfns}

The five-dimensional Kerr--(A)dS solution was constructed in \cite{Hawking:1998kw} and
generalized to arbitrary spacetime dimension in \cite{Gibbons:2004uw}. It has a compact
presentation in Kerr--Schild form, but the separability we are after is visible only in
Boyer--Lindquist coordinates \( x^{\mu} = (\tau,\, r,\, \theta,\, \ph_1,\, \ph_2) \), in
which the metric reads
\begin{equation} \label{eq:ds2}
\begin{split}
  \dd s^{2}
  ={}& -\,W \Xir \,\dd\tau^{2}
     \;+\; \frac{\rhs\, \dd r^{2}}{V-2M}
     \;+\; \frac{\rhs}{\Dth}\,\dd\theta^{2} \\[4pt]
     & +\; \frac{2M}{\rhs}
       \left( \dd\tau - \sum_{i=1}^{2}
              \frac{a_i \mu_i^{2}\,\dd\ph_i}{\Xii} \right)^{\!\!2}
     \;+\; \sum_{i=1}^{2} \frac{\Ri}{\Xii}\,\mu_i^{2}\,
           \bigl( \dd\ph_i - \lambda a_i \,\dd\tau \bigr)^{2} \,.
\end{split}
\end{equation}
Here $M$ is the mass parameter, \( a_1 := a \) and \( a_2 := b \) are the two independent
rotation parameters, and $\lambda$ is proportional to the cosmological constant. We write
\( \mu_1 = \sin\theta \) and \( \mu_2 = \cos\theta \), so that \( \mu_1^2 + \mu_2^2 = 1 \).
Indices $i$ and $j$ take the values $1$ and $2$, and repeated ones are summed wherever no
explicit summation sign appears. The azimuthal coordinates $\ph_i$ are periodic with period
$2\pi$, whereas \( \theta \in [0,\pi/2] \), the two endpoints being the axes of the two
rotation planes. Following \cite{Kunduri:2005fq} we abbreviate
\begin{equation} \label{eq:abbrev}
  \Xii := 1 + \lambda a_i^{2} \,, \qquad
  \Xir := 1 - \lambda r^{2} \,, \qquad
  \Ri  := r^{2} + a_i^{2} \,,
\end{equation}
so that \( \Xia := \Xi_1 = 1+\lambda a^{2} \), \( \Xib := \Xi_2 = 1+\lambda b^{2} \),
\( R_1 = r^{2} + a^{2} \) and \( R_2 = r^{2} + b^{2} \). The remaining metric functions
appearing in \eqref{eq:ds2} are
\begin{align}
  \chit  &:= a^{2}\mu_2^{2} + b^{2}\mu_1^{2}
             = a^{2}\cos^{2}\theta + b^{2}\sin^{2}\theta ,
  \label{eq:chi}\\[2pt]
  \rhs   &:= r^{2} + \chit \,,
  \label{eq:rho2}\\[2pt]
  \Dth   &:= 1 + \lambda \chit \,,
  \label{eq:Dth}\\[2pt]
  V      &:= \frac{\Xir\, R_1 R_2}{r^{2}}
          \;=\; \frac{\Xir\,(r^{2}+a^{2})(r^{2}+b^{2})}{r^{2}} ,
  \label{eq:V}\\[2pt]
  W      &:= \sum_{i=1}^{2}\frac{\mu_i^{2}}{\Xii}
          \;=\; \frac{\Dth}{\Xia \Xib} .
  \label{eq:W}
\end{align}
The polar variable $\chit$ of \eqref{eq:chi} is introduced because it makes the pairing
\begin{equation} \label{eq:pairing}
  \Xir = 1 - \lambda r^{2} \,, \qquad \Dth = 1 + \lambda \chit \,, \qquad
  \rhs = r^{2} + \chit \,, \qquad \Dth - \Xir = \lambda\rhs \,,
\end{equation}
manifest. It is this pairing that controls the separation in
Sec.~\ref{sec:hj}. The five-dimensional Kerr--(A)dS metric \eqref{eq:ds2} satisfies
\( R_{\mu\nu} = 4\lambda g_{\mu\nu} \), and for generic values of the rotation parameters
$a$ and $b$, it admits the three commuting Killing vectors
\( \partial_\tau, \partial_{\ph_1}, \partial_{\ph_2} \). The volume element for this
coordinate chart is
\begin{equation} \label{eq:sqrtg}
  \sqrt{-g} = \frac{r\,\rhs\,\sin\theta \, \cos\theta}{\Xia \Xib} \,,
\end{equation}
which is positive throughout \( 0 < \theta < \pi/2 \) and vanishes on the two axes.

%%-------------------------------------------------------------------- II.B --
\subsection{The inverse metric}\label{sec:inverse}

What the Hamilton--Jacobi equation needs is $g^{\mu\nu}$, and in a form in which its $r$
and $\theta$ dependence can be told apart. Two ingredients bring it to that form. The first
is the combination of metric functions
\begin{equation} \label{eq:kappa}
   \kap := \frac{2MV}{\rhs\,(V-2M)} \;=\; \frac{2M}{\rhs} + \frac{4M^{2}}{\rhs\,(V-2M)} \,,
\end{equation}
and the second is the identity
\begin{equation} \label{eq:key}
  \lambda \Ri + \Xir
  = \lambda\bigl(r^{2}+a_i^{2}\bigr) + 1 - \lambda r^{2}
  = 1 + \lambda a_i^{2}
  = \Xii \,,
  \qquad \mbox{i.e.} \qquad
  \Xii - \Xir = \lambda \Ri \,,
\end{equation}
which relates the radial and the rotational abbreviations of \eqref{eq:abbrev} and will be
used again and again below. With these two, the nontrivial components of the inverse metric
read
\begin{subequations} \label{eq:inv}
\begin{align}
  g^{rr} &= \frac{V-2M}{\rhs} \,,
  \qquad
  g^{\theta\theta} = \frac{\Dth}{\rhs} \,,
  \label{eq:invrr}\\[4pt]
  g^{\tau\tau}
    &= -\frac{1}{W \Xir} - \frac{\kap}{\Xir^{2}} \,,
  \label{eq:invtt}\\[4pt]
  g^{\tau\ph_i}
    &= \lambda a_i\, g^{\tau\tau}
       \;-\; \frac{\kap\, a_i}{\Xir \Ri} \,,
  \label{eq:invtp}\\[4pt]
  g^{\ph_i\ph_j}
    &= \frac{\Xii\,\delta^{ij}}{\Ri\,\mu_i^{2}}
       \;+\; \lambda^{2} a_i a_j\, g^{\tau\tau}
       \;-\; \kap\, a_i a_j
         \left[ \frac{\lambda}{\Xir}
                \left( \frac{1}{\Ri} + \frac{1}{\Rj} \right)
              + \frac{1}{\Ri \Rj} \right] .
  \label{eq:invpp}
\end{align}
\end{subequations}
Note that the functions $\rhs\, g^{\mu\nu}$ built from \eqref{eq:inv} are additively
separable as functions of $r$ and $\theta$, which is the structural feature exploited in
the Hamilton--Jacobi analysis of Sec.~\ref{sec:hj}.

%%-------------------------------------------------------------------- II.C --
\subsection{Comparison with Kunduri and Lucietti, and a correction}\label{sec:compare}

The inverse metric is also given in \cite{Kunduri:2005fq}. Two of its three blocks agree
with \eqref{eq:inv} and the third does not, so we set them side by side. To compare
\eqref{eq:inv} with Eqs.~(6)--(8) of \cite{Kunduri:2005fq}, one can expand
$\kap$ using \eqref{eq:kappa} and introduce
\begin{equation} \label{eq:Q}
  Q := -\frac{1}{W\Xir} - \frac{2M}{\rhs\,\Xir^{2}} \,,
\end{equation}
of \cite{Kunduri:2005fq}, so that
\begin{equation} \label{eq:gttpaper}
  g^{\tau\tau} = Q - \frac{4M^{2}}{\rhs\,\Xir^{2}\,(V-2M)} \,,
\end{equation}
in agreement with \cite{Kunduri:2005fq}. For the mixed component, we find
\begin{equation} \label{eq:gtpi}
  g^{\tau\ph_i} =  \lambda a_i Q
     - \frac{2M a_i}{\rhs\, \Xir \Ri}
     - \frac{4M^{2} a_i \Xii}{\rhs\, \Xir^{2}\,(V-2M)\, \Ri} \,,
\end{equation}
which also agrees with \cite{Kunduri:2005fq}. However, for the final bit we find
\begin{equation} \label{eq:invpp-paperstyle}
  g^{\ph_i\ph_j}
  = \frac{\Xii\,\delta^{ij}}{\Ri\,\mu_i^{2}}
    + \lambda^{2} a_i a_j\, Q
    + Q^{ij} \,,
\end{equation}
where
\begin{equation} \label{eq:Qij}
  Q^{ij}
  := -\frac{2M a_i a_j}{\rhs\, \Ri \Rj}
    - \frac{2M\lambda a_i a_j}{\rhs\, \Xir}
      \left[ \frac{1}{\Ri} + \frac{1}{\Rj} \right]
    - \frac{4M^{2} a_i a_j \Xii \Xij}
           {\rhs\, \Xir^{2}\,(V-2M)\, \Ri \Rj} \,,
\end{equation}
which, unfortunately, does not agree with \cite{Kunduri:2005fq}.

No erratum to \cite{Kunduri:2005fq} appears to have been published, nor, as far as we have
been able to determine, has this point been remarked upon in the literature citing it. The
uncharged limit of the inverse metric given in \cite{Davis:2005ys} agrees with the
corrected $g^{\ph_i\ph_j}$ block \eqref{eq:invpp-paperstyle} obtained here, rather than
with the one published in \cite{Kunduri:2005fq}. Our result \eqref{eq:inv} likewise agrees
with \cite{Wu:2008bc}, where the inverse metric is written as a sum of squares of dual
one-forms rather than in components. The error has, however, propagated to general $D$ in
\cite{Vasudevan:2005js}, where it also enters the radial geodesic equation. The same block
is given incorrectly, in a manner different again, in \cite{Li:2009zzs}.

%%%%%%%%%%%%%%%%%%%%%%%%%%%%%%%%%%%%%%%%%%%%%%%%%%%%%%%%%%%%%%%%%%%%%%%%%%%%%%
%%  III.  SEPARABILITY OF THE HAMILTON-JACOBI EQUATION
%%%%%%%%%%%%%%%%%%%%%%%%%%%%%%%%%%%%%%%%%%%%%%%%%%%%%%%%%%%%%%%%%%%%%%%%%%%%%%
\section{Separability of the Hamilton--Jacobi equation}\label{sec:hj}

Separability of the Hamilton--Jacobi equation for the five-dimensional Myers--Perry black
hole, that is the $\lambda = 0$ case, was established in \cite{Frolov:2003en}, and with a
cosmological constant in \cite{Kunduri:2005fq}. Nothing in this section or the next is new
as physics: the separated equations are known, and in the form given below they agree with
the independent derivation of \cite{Grunau:2017uwv}, as we verify in
Sec.~\ref{sec:earlier}. What we do is to correct them where they are given incorrectly in
\cite{Kunduri:2005fq}, to normalize the separation constant so that the Myers--Perry limit
survives, and to carry timelike and null geodesics together, so that the first-order system
is in the shape required by the Biermann--Weierstrass construction of
Sec.~\ref{sec:bw}. The three subsections below set up the problem, isolate the two
identities on which the splitting turns, and record the separated equations.

%%------------------------------------------------------------------- III.A --
\subsection{Setup and the generating function}\label{sec:setup}

We begin by fixing the variational setup and the generating function, going carefully over
the discussion given in Eqs.~(10)--(16) of \cite{Kunduri:2005fq}.
Consider a free particle moving on a geodesic with affine parameter $l$, and set
\begin{equation} \label{eq:pdown}
 p_{\mu} = g_{\mu\nu} \frac{\dd x^{\nu}}{\dd l} \,,
 \qquad
 g_{\mu\nu} \frac{\dd x^{\mu}}{\dd l}\frac{\dd x^{\nu}}{\dd l} = \eps \,,
 \qquad
 \eps = \begin{cases} -1 & \mbox{timelike geodesics,} \\[2pt]
                      \phantom{-}0 & \mbox{null geodesics.} \end{cases}
\end{equation}
For $\eps = -1$ the parameter $l$ is proper time, $p_\mu$ is the momentum per unit mass, and
the momentum of a particle of mass $m$ is $m\,p_\mu$. For $\eps = 0$ the parameter $l$ is any
affine parameter along the ray. Reference \cite{Kunduri:2005fq} treats only the timelike
case, writing $s = m\,l$ and normalizing to $-m^{2}$ instead of $\eps$. Its equations are
recovered from ours by $\eps \to -m^{2}$, with $E$, $L_i$ and $K$ then carrying the
corresponding factors of $m$. The Hamiltonian
\( H = \half\, g^{\mu\nu} p_{\mu} p_{\nu} \) follows easily. It carries no explicit
$l$-dependence, and is conserved with the on-shell value \( H = \eps/2 \).
So now one looks for a type-II generating function $S(x^{\mu}, P_{\nu}, l)$ for a canonical
transformation \( (x^{\mu}, p_{\nu}) \to (X^{\mu}, P_{\nu}) \) to \emph{constant} new variables.
One must have
\begin{equation} \label{eq:typeII}
  p_{\mu} = \frac{\partial S}{\partial x^{\mu}}\,, \qquad
  X^{\mu} = \frac{\partial S}{\partial P_{\mu}}\,, \qquad
  C = H + \frac{\partial S}{\partial l}\,,
\end{equation}
where $C$ is the new Hamiltonian. Demanding $C=0$ gives Eq.~(10) of \cite{Kunduri:2005fq}:
\begin{equation} \label{eq:Seqn}
  \frac{\partial S}{\partial l} + \frac{1}{2}\, g^{\mu\nu}
  \frac{\partial S}{\partial x^{\mu}} \frac{\partial S}{\partial x^{\nu}} = 0 \,.
\end{equation}
From \eqref{eq:Seqn}, \( \partial_{l} S = - H = -\eps/2 = \) const., and using the three
Killing vectors one can write
\begin{equation} \label{eq:Ssol}
  S = -\frac{1}{2}\, \eps\, l - E \tau + \sum_{i=1}^{2} L_i \ph_i + F(r, \theta) \,,
\end{equation}
with the old momenta \( p_\mu = (-E, \partial_r F, \partial_\theta F, L_1, L_2) \). The
function $F(r, \theta)$ is an arbitrary function of the two non-cyclic coordinates at this
stage. With the complete set of new momenta\footnote{See \eqref{eq:theta-eq} for the
separation constant $K$.} \( P_{\mu} = (\eps, E, L_1, L_2, K) \), one has five constants for
a five-dimensional configuration space, which is necessary for Liouville integrability.

%%------------------------------------------------------------------- III.B --
\subsection{Two identities and the choice of separation constant}\label{sec:ids}

Everything now turns on two identities, and on a choice that comes with the first of them.
What they are for is the splitting of $F(r, \theta)$ as \( S_r(r) + S_\theta(\theta) \).
The first concerns the single non-separable-looking structure in the constraint, namely
$\rhs/(W\Xir) = \Xia\Xib\,\rhs/(\Xir\Dth)$, which must be written as a function of $r$ plus
a function of $\theta$. Using $\rhs = r^{2}+\chit$ together with $\Dth = 1+\lambda\chit$ and
$\Xir = 1-\lambda r^{2}$ one has $r^{2}\Dth + \chit\,\Xir = \rhs$, that is
\begin{equation} \label{eq:id12}
  \frac{\rhs}{W\Xir}
  = \Xia\Xib \left[ \frac{r^{2}}{\Xir} + \frac{\chit}{\Dth} \right] .
\end{equation}
The second,
\begin{equation} \label{eq:id13}
  \sum_{i=1}^{2} \frac{L_i^{2}\Xii}{\mu_i^{2}\Ri}\,\rhs
  = \sum_{i=1}^{2} \frac{L_i^{2}\Xii}{\mu_i^{2}}
    + \frac{L_1^{2}\Xia\,(b^{2}-a^{2})}{R_1}
    + \frac{L_2^{2}\Xib\,(a^{2}-b^{2})}{R_2} \,,
\end{equation}
is Eq.~(13) of \cite{Kunduri:2005fq} and follows from
$\rhs = \Ri + (a_{j}^{2}-a_i^{2})\mu_i^{2}$ with $j \neq i$.

A splitting of the kind \eqref{eq:id12} is never unique: any constant may be added to the
radial half and subtracted from the polar half, and the separation constant changes
accordingly. In place of \eqref{eq:id12}, Ref.~\cite{Kunduri:2005fq} uses the relation
$\Dth - \Xir = \lambda \rhs$ in its Eq.~(12) to write
\begin{equation} \label{eq:id12KL}
  \frac{\rhs}{W\Xir}
  = \frac{\Xia\Xib}{\lambda} \left[ \frac{1}{\Xir} - \frac{1}{\Dth} \right] .
\end{equation}
The two are the same identity: \eqref{eq:id12KL} is obtained from \eqref{eq:id12} by moving
the constant $\Xia\Xib/\lambda$ from the polar half to the radial half, since
$1/(\lambda\Xir) - r^{2}/\Xir = 1/\lambda$ and
$\chit/\Dth + 1/(\lambda\Dth) = 1/\lambda$. They are not, however, equally convenient. The
choice \eqref{eq:id12KL} introduces a pole at $\lambda = 0$ into \emph{each} of the two
separated equations, and these poles cancel only when the equations are recombined. The
choice \eqref{eq:id12} introduces none. We therefore adopt \eqref{eq:id12} throughout. The price
is that our separation constant differs from the one used in \cite{Kunduri:2005fq} by
exactly the constant that was moved,
\begin{equation} \label{eq:KvsKKL}
  K = \Kkl - \frac{\Xia\Xib\,\Ecal^{2}}{\lambda} \,,
\end{equation}
where $\Ecal := E - \lambda a_i L_i$ is the co-rotating energy of \eqref{eq:constraint}
below. The benefit is that every equation from here on is regular at $\lambda = 0$, and the
Myers--Perry limit taken at the end of Sec.~\ref{sec:tauphi} is a plain substitution rather
than a shift performed after the fact.

%%------------------------------------------------------------------- III.C --
\subsection{The separated equations}\label{sec:separated}

We can now carry out the separation and record the two ordinary differential equations it
produces, together with their relation to the equations as published.
Using the explicit formulas \eqref{eq:inv} and the co-rotating energy
\( \Ecal := E - \lambda a_i L_i = E - \lambda\,( a L_1 + b L_2) \), we find
\begin{equation} \label{eq:constraint}
  g^{\mu\nu} p_\mu p_\nu
  = -\frac{\Ecal^{2}}{W \Xir}
    + \sum_{i=1}^{2}\frac{\Xii L_i^{2}}{\Ri \mu_i^{2}}
    + \frac{V-2M}{\rhs}\,p_r^{2}
    + \frac{\Dth}{\rhs}\,p_\theta^{2}
    - \kap\,\Psi^{2} \,,
\end{equation}
where
\begin{equation} \label{eq:Psi}
  \Psi(r) := \frac{\Ecal}{\Xir} - \sum_{i=1}^{2} \frac{a_i L_i}{\Ri}
\end{equation}
depends on $r$ alone. Since $\kap\Psi^{2} = 2MV\Psi^{2}/[\rhs(V-2M)]$, the entire
$M$-dependence of \eqref{eq:constraint} is $1/\rhs$ times a function of $r$, and separability
follows: multiplying by $\rhs$ and using \eqref{eq:id12}--\eqref{eq:id13} splits the
normalization condition $g^{\mu\nu}p_\mu p_\nu = \eps$ into a function of $\theta$ plus a
function of $r$, each of which must therefore be constant. The $\theta$ equation is
\begin{equation} \label{eq:theta-eq}
  \Dth \left(\frac{\dd S_\theta}{\dd \theta}\right)^{\!2}
  - \left( \eps + \frac{\Xia\Xib\,\Ecal^{2}}{\Dth} \right) \chit
  + \frac{\Xia L_1^{2}}{\sin^{2}\theta}
  + \frac{\Xib L_2^{2}}{\cos^{2}\theta}
  = K \,,
\end{equation}
$K$ being the separation constant, and the radial equation is
\begin{equation} \label{eq:r-eq}
  (V-2M)\left(\frac{\dd S_r}{\dd r}\right)^{\!2}
  + \widetilde{V}(r; E, L_i, \eps) = -K \,,
\end{equation}
with the effective potential
\begin{equation} \label{eq:Vtilde}
  \widetilde{V}(r)
  = -\left( \eps + \frac{\Xia\Xib\, \Ecal^{2}}{\Xir} \right) r^{2}
    + \frac{\Xia (b^{2}-a^{2}) L_1^{2}}{R_1}
    + \frac{\Xib (a^{2}-b^{2}) L_2^{2}}{R_2}
    - \frac{2MV}{V-2M}\,\Psi^{2} \,,
\end{equation}
manifestly a function of $r$ alone. The whole of its $M$-dependence has been collected into
the single square $-2MV\Psi^{2}/(V-2M)$. The parallel between \eqref{eq:theta-eq} and
\eqref{eq:Vtilde} is the pairing \eqref{eq:pairing} at work: the factor
$\eps + \Xia\Xib\Ecal^{2}/\Xir$ multiplies $r^{2}$ in the radial equation exactly as
$\eps + \Xia\Xib\Ecal^{2}/\Dth$ multiplies $\chit$ in the polar one. Neither equation
contains a pole at $\lambda = 0$, and neither requires $\eps \neq 0$: null geodesics are
covered throughout by setting $\eps = 0$.

Equations \eqref{eq:theta-eq} and \eqref{eq:r-eq} are Eqs.~(14) and (15) of
\cite{Kunduri:2005fq} rewritten with the separation constant \eqref{eq:KvsKKL}: adding
$\Xia\Xib\Ecal^{2}/\lambda$ to both sides of \eqref{eq:theta-eq} turns its left-hand side
into that of their Eq.~(14), because
$\Xia\Xib\Ecal^{2}\bigl(\lambda^{-1} - \chit/\Dth\bigr) = \Xia\Xib\Ecal^{2}/(\lambda\Dth)$
by $\Dth - \lambda\chit = 1$, while its right-hand side becomes $\Kkl$. The same holds for
\eqref{eq:r-eq}, with $\Xir + \lambda r^{2} = 1$. Denoting by $\widetilde{V}_{\rm KL}$ the
potential obtained from Eq.~(16)
of \cite{Kunduri:2005fq}, shifted by the same constant so that the two are compared at
equal $K$, the discrepancy inherited from $Q^{ij}$ is
\begin{equation} \label{eq:Vexcess}
  \widetilde{V}_{\rm KL} - \widetilde{V}
  = \frac{8M^{2}\,(1+\Xir)}{\Xir^{2}\,(V-2M)}
    \left( \sum_{i=1}^{2}\frac{\Xii a_i L_i}{\Ri} \right)
    \left( \sum_{j=1}^{2}\frac{a_j L_j}{\Rj} \right) ,
\end{equation}
which is $O(M^{2})$ but survives the limit $\lambda \to 0$. It vanishes when $aL_1$ and
$bL_2$ both vanish, and, for $a=b$, also when $L_2 = -L_1$. The vanishing of the single
combination $a_i L_i$ is not enough, since each sum in \eqref{eq:Vexcess} is weighted by
$1/\Ri$.\footnote{In Appendix \ref{app:kg} we briefly discuss the effects of the
error we have identified on the separability of the Klein--Gordon equation examined in
\cite{Kunduri:2005fq}.}

%%%%%%%%%%%%%%%%%%%%%%%%%%%%%%%%%%%%%%%%%%%%%%%%%%%%%%%%%%%%%%%%%%%%%%%%%%%%%%
%%  IV.  FIRST-ORDER FORM OF THE GEODESIC EQUATIONS
%%%%%%%%%%%%%%%%%%%%%%%%%%%%%%%%%%%%%%%%%%%%%%%%%%%%%%%%%%%%%%%%%%%%%%%%%%%%%%
\section{First-order form of the geodesic equations}\label{sec:firstorder}

The separation of Sec.~\ref{sec:hj} determines the momenta. This section turns them into a
first-order system for the coordinates themselves, decouples the radial from the polar
motion by a change of parameter, removes the overall scale, and ends with the two
polynomials that the whole of Sec.~\ref{sec:bw} rests on.

Raising the index on \eqref{eq:pdown} gives, with no further input,
\begin{equation} \label{eq:pup}
  p^{\mu} \;=\; g^{\mu\nu}p_\nu \;=\; \frac{\dd x^{\mu}}{\dd l} \,,
\end{equation}
so that the contravariant momentum is the tangent vector itself. Combining \eqref{eq:pup}
with $p_\mu = \partial_\mu S$ turns each component into a first-order ordinary differential
equation. Explicitly, since the
metric carries no off-diagonal component involving $\dd r$ or $\dd\theta$,
\begin{equation} \label{eq:rtheta-first}
  \frac{\dd r}{\dd l} = g^{rr}p_r
    = g^{rr}\frac{\dd S_r}{\dd r} \,,
  \qquad
  \frac{\dd\theta}{\dd l} = g^{\theta\theta}p_\theta
    = g^{\theta\theta}\frac{\dd S_\theta}{\dd\theta} \,,
\end{equation}
while the two Killing directions mix,
\begin{equation} \label{eq:tauphi-first}
  \frac{\dd\tau}{\dd l} = -E\,g^{\tau\tau} + L_i\,g^{\tau\ph_i} \,,
  \qquad
  \frac{\dd\ph_i}{\dd l} = -E\,g^{\tau\ph_i} + L_j\,g^{\ph_i\ph_j} \,.
\end{equation}

%%-------------------------------------------------------------------- IV.A --
\subsection{Radial and polar equations}\label{sec:radpol}

The radial and polar equations come first, because they close among themselves.
From \eqref{eq:rtheta-first} with \eqref{eq:invrr}, and then using \eqref{eq:r-eq} and
\eqref{eq:theta-eq} to eliminate the derivatives of $S_r$ and $S_\theta$,
\begin{equation} \label{eq:drdtheta}
  \rhs\,\frac{\dd r}{\dd l} = \pm\sqrt{\Rcal(r)} \,,
  \qquad
  \rhs\,\frac{\dd\theta}{\dd l} = \pm\sqrt{\Tcal(\theta)} \,,
\end{equation}
where
\begin{equation} \label{eq:RTdef}
\begin{split}
  \Rcal(r) &:= (V-2M)\bigl[\, -K - \widetilde{V}(r) \,\bigr] \,, \\[4pt]
  \Tcal(\theta) &:= \Dth
    \left[ K + \eps\,\chit
         - \frac{\Xia L_1^{2}}{\sin^{2}\theta}
         - \frac{\Xib L_2^{2}}{\cos^{2}\theta} \right]
       + \Xia\Xib\,\Ecal^{2}\,\chit \,.
\end{split}
\end{equation}
The two signs are independent and flip at the turning points $\Rcal=0$ and $\Tcal=0$.

%%-------------------------------------------------------------------- IV.B --
\subsection{Temporal and azimuthal equations}\label{sec:tauphi}

The remaining three coordinates are cyclic, and their equations mix through the Killing
block of the inverse metric. Collecting the terms around one radial function keeps that
mixing manageable. It is convenient to define
\begin{equation} \label{eq:kahat}
   \kah(r) := \rhs\,\kap = \frac{2MV}{V-2M} \,,
\end{equation}
which depends on $r$ alone. Substituting \eqref{eq:inv} into the first of
\eqref{eq:tauphi-first} and collecting the terms proportional to $g^{\tau\tau}$,
\begin{equation} \label{eq:dtau-compact}
  \frac{\dd\tau}{\dd l}
  = -\Ecal\, g^{\tau\tau} - \frac{\kap}{\Xir}\sum_{i=1}^{2} \frac{a_i L_i}{\Ri}
  = \frac{\Ecal}{W\Xir}
    + \frac{\kap}{\Xir}
      \left[ \frac{\Ecal}{\Xir} - \sum_{i=1}^{2} \frac{a_i L_i}{\Ri} \right]
  = \frac{\Ecal}{W\Xir} + \frac{\kap\,\Psi}{\Xir} \,,
\end{equation}
the bracket assembling into precisely the function $\Psi$ of \eqref{eq:Psi}. Multiplying by
$\rhs$ and using \eqref{eq:id12},
\begin{equation} \label{eq:dtau}
   \rhs\,\frac{\dd\tau}{\dd l}
  = \Xia\Xib\,\Ecal
      \left( \frac{r^{2}}{\Xir} + \frac{\chit}{\Dth} \right)
    + \frac{\kah(r)\,\Psi(r)}{\Xir} \,.
\end{equation}
The same substitution in the second of \eqref{eq:tauphi-first} gives, for each $i$ and with
no sum on $i$,
\begin{equation} \label{eq:dphi-compact}
  \frac{\dd\ph_i}{\dd l}
  = \frac{\Xii L_i}{\Ri\,\mu_i^{2}}
    + \frac{\lambda a_i \Ecal}{W\Xir}
    + \kap\, a_i \Psi
      \left[ \frac{\lambda}{\Xir} + \frac{1}{\Ri} \right]
  = \frac{\Xii L_i}{\Ri\,\mu_i^{2}}
    + \frac{\lambda a_i \Ecal}{W\Xir}
    + \frac{\kap\, a_i \Xii \Psi}{\Xir \Ri} \,,
\end{equation}
where the last step is again the identity \eqref{eq:key}. Multiplying by $\rhs$ and using
\eqref{eq:id12} together with
$\rhs/\Ri = 1 + (a_j^{2}-a_i^{2})\mu_i^{2}/\Ri$ \ ($j \neq i$),
\begin{equation} \label{eq:dphi}
  \rhs\,\frac{\dd\ph_i}{\dd l}
  = \frac{\Xii L_i}{\mu_i^{2}}
    + \frac{\Xii\,(a_j^{2}-a_i^{2})\,L_i}{\Ri}
    + \lambda a_i \Xia\Xib\, \Ecal
      \left( \frac{r^{2}}{\Xir} + \frac{\chit}{\Dth} \right)
    + \frac{\kah(r)\, a_i \Xii \Psi(r)}{\Xir \Ri} \,,
\end{equation}
again with $j \neq i$ and no sum on $i$. Equations
\eqref{eq:drdtheta}, \eqref{eq:dtau} and \eqref{eq:dphi} are the announced first-order
system. Note that every right-hand side is a sum of a function of $r$ and a function of
$\theta$: in \eqref{eq:dtau} and \eqref{eq:dphi} the $r^{2}/\Xir$ terms are radial and the
$\chit/\Dth$ terms polar.

The separation constant was normalized in Sec.~\ref{sec:ids} so as to keep the splitting
identity \eqref{eq:id12} free of poles. The Myers--Perry limit of the system
\eqref{eq:drdtheta}, \eqref{eq:dtau} and \eqref{eq:dphi} is therefore taken by simply
setting $\lambda = 0$. One has $\Xii, \Xir, \Dth \to 1$ and $\Ecal \to E$, the two halves of
\eqref{eq:id12} recombine into $r^{2} + \chit = \rhs$ in \eqref{eq:dtau}, and the
corresponding term of \eqref{eq:dphi} drops out because it carries an explicit factor of
$\lambda$. Written in the variable $x = r^{2}$, with $\eps \to -m^{2}$ and with the
orientation of the rotation planes reversed, the result is precisely Eqs.~(3.26)--(3.30) of
\cite{Frolov:2003en}.

%%-------------------------------------------------------------------- IV.C --
\subsection{Reparametrization by the Mino parameter}\label{sec:repara}

The radial and polar equations \eqref{eq:drdtheta} are coupled only through the common
factor $\rhs$ on their left-hand sides. Following \cite{Mino:2003yg}, that coupling is
removed by passing to a new parameter $\zeta$ defined by
\begin{equation} \label{eq:mino}
  \dd\zeta := \frac{\dd l}{\rhs} \,,
  \qquad \mbox{equivalently} \qquad
  l = \int_{0}^{\zeta}
      \rhs\bigl( r(\tilde\zeta),\, \theta(\tilde\zeta) \bigr) \,\dd\tilde\zeta \,,
\end{equation}
in terms of which the system \eqref{eq:drdtheta}, \eqref{eq:dtau} and \eqref{eq:dphi}
becomes
\begin{align}
  \frac{\dd r}{\dd\zeta} &= \pm\sqrt{\Rcal(r)} \,,
  \qquad
  \frac{\dd\theta}{\dd\zeta} = \pm\sqrt{\Tcal(\theta)} \,,
  \label{eq:zeta-rtheta}\\[4pt]
  \frac{\dd\tau}{\dd\zeta}
    &= \Xia\Xib\,\Ecal
       \left( \frac{r^{2}}{\Xir} + \frac{\chit}{\Dth} \right)
     + \frac{\kah(r)\,\Psi(r)}{\Xir} \,,
  \label{eq:zeta-tau}\\[4pt]
  \frac{\dd\ph_i}{\dd\zeta}
    &= \frac{\Xii L_i}{\mu_i^{2}}
     + \frac{\Xii\,(a_j^{2}-a_i^{2})\,L_i}{\Ri}
     + \lambda a_i \Xia\Xib\, \Ecal
       \left( \frac{r^{2}}{\Xir} + \frac{\chit}{\Dth} \right)
     + \frac{\kah(r)\, a_i \Xii \Psi(r)}{\Xir \Ri} \,,
  \label{eq:zeta-phi}
\end{align}
again with $j \neq i$ and no sum on $i$. The two equations \eqref{eq:zeta-rtheta} are now
completely decoupled, each involving a single variable, and once $r(\zeta)$ and
$\theta(\zeta)$ are known the remaining three follow by quadrature. That one
reparametrization suffices is particular to four and five dimensions, a point we return to
in Sec.~\ref{sec:conclusions}.

%%-------------------------------------------------------------------- IV.D --
\subsection{Dimensionless form of the system}\label{sec:rescaled}

One of the nine parameters of the system is pure scale and can be removed. In five
dimensions the mass parameter carries the dimensions of a squared length, so that
\begin{equation} \label{eq:r0}
  r_0 := \sqrt{2M}
\end{equation}
is the natural length scale of the problem. Indeed
\eqref{eq:zeta-rtheta}--\eqref{eq:zeta-phi} are invariant under
\begin{equation} \label{eq:scaling}
\begin{split}
  & r \to c\,r \,, \qquad a_i \to c\,a_i \,, \qquad \tau \to c\,\tau \,, \qquad
    l \to c\,l \,, \qquad \zeta \to \zeta/c \,, \\[4pt]
  & \lambda \to \lambda/c^{2} \,, \qquad M \to c^{2}M \,, \qquad
    L_i \to c\,L_i \,, \qquad K \to c^{2}K \,,
\end{split}
\end{equation}
at fixed $\theta$, $\ph_i$, $E$ and $\eps$. Following the treatment of the Kerr case in
Sec.~II of \cite{Cieslik:2023qdc}, we use this freedom to pass to dimensionless variables.
Setting $c = 1/r_0$, write
\begin{align}
  r &= r_0\,\xi \,, \qquad a = r_0\,\alpha \,, \qquad b = r_0\,\beta \,, \qquad
     \tau = r_0\,T \,, \qquad l = r_0\,\eta \,, \qquad \zeta = \gamma/r_0 \,,
  \label{eq:rescale1}\\[2pt]
  \lambda &= \frac{\Lambda}{r_0^{2}} \,, \qquad
     L_i = r_0\,\ell_i \,, \qquad
     K = r_0^{2}\,\Kcal \,,
  \label{eq:rescale2}
\end{align}
the energy $E$ and the normalization $\eps$ being dimensionless already. The structure
functions of Sec.~\ref{sec:metricfns} that are themselves dimensionless keep their form
verbatim,
\begin{equation} \label{eq:barinvariant}
  \Xia = 1 + \Lambda\alpha^{2} \,, \quad
  \Xib = 1 + \Lambda\beta^{2} \,, \quad
  \Xir = 1 - \Lambda\xi^{2} \,, \quad
  \Dth = 1 + \Lambda\bar\chi \,, \quad
  W = \frac{\Dth}{\Xia\Xib} \,,
\end{equation}
while those of dimension length squared acquire one factor of $r_0^{2}$. Denoting the
rescaled quantity by a bar,
\begin{equation} \label{eq:barred}
  \bar\chi := \alpha^{2}\cos^{2}\theta + \beta^{2}\sin^{2}\theta \,, \qquad
  \bar\rho^{2} := \xi^{2} + \bar\chi \,, \qquad
  \bar R_i := \xi^{2} + \alpha_i^{2} \,, \qquad
  \bar V := \frac{\Xir\,\bar R_1 \bar R_2}{\xi^{2}} \,,
\end{equation}
with $\alpha_1 := \alpha$ and $\alpha_2 := \beta$. The combination $V - 2M$ becomes
$r_0^{2}(\bar V - 1)$, which is where the choice \eqref{eq:r0} pays off, and correspondingly
\begin{equation} \label{eq:barkappa}
  \Ecal = E - \Lambda\,\alpha_i \ell_i \,, \qquad
  \Psi = \frac{\Ecal}{\Xir} - \sum_{i=1}^{2}\frac{\alpha_i \ell_i}{\bar R_i} \,, \qquad
  \bar\kappa := \frac{\kah}{r_0^{2}} = \frac{\bar V}{\bar V - 1} \,,
\end{equation}
so that $\Ecal$ and $\Psi$, being dimensionless, are also unchanged in form. Writing
finally $\bar V_{\!\ast} := \widetilde{V}/r_0^{2}$, $\bar\Rcal := \Rcal/r_0^{4}$ and
$\bar\Tcal := \Tcal/r_0^{2}$, that is
\begin{align}
  \bar V_{\!\ast} &= -\left( \eps + \frac{\Xia\Xib\,\Ecal^{2}}{\Xir} \right) \xi^{2}
     + \frac{\Xia(\beta^{2}-\alpha^{2})\,\ell_1^{2}}{\bar R_1}
     + \frac{\Xib(\alpha^{2}-\beta^{2})\,\ell_2^{2}}{\bar R_2}
     - \bar\kappa\,\Psi^{2} \,,
  \label{eq:barVeff}\\[4pt]
  \bar\Rcal &= (\bar V - 1)\bigl[\, -\Kcal - \bar V_{\!\ast} \,\bigr] \,,
  \label{eq:barR}\\[4pt]
  \bar\Tcal &= \Dth \left[ \Kcal + \eps\,\bar\chi
     - \frac{\Xia \ell_1^{2}}{\sin^{2}\theta}
     - \frac{\Xib \ell_2^{2}}{\cos^{2}\theta} \right]
     + \Xia\Xib\,\Ecal^{2}\,\bar\chi \,,
  \label{eq:barT}
\end{align}
the system \eqref{eq:zeta-rtheta}--\eqref{eq:zeta-phi} takes exactly the same form in the
rescaled variables,
\begin{align}
  \frac{\dd\xi}{\dd\gamma} &= \pm\sqrt{\bar\Rcal(\xi)} \,,
  \qquad
  \frac{\dd\theta}{\dd\gamma} = \pm\sqrt{\bar\Tcal(\theta)} \,,
  \label{eq:bar-rtheta}\\[4pt]
  \frac{\dd T}{\dd\gamma}
    &= \Xia\Xib\,\Ecal
       \left( \frac{\xi^{2}}{\Xir} + \frac{\bar\chi}{\Dth} \right)
     + \frac{\bar\kappa(\xi)\,\Psi(\xi)}{\Xir} \,,
  \label{eq:bar-tau}\\[4pt]
  \frac{\dd\ph_i}{\dd\gamma}
    &= \frac{\Xii \ell_i}{\mu_i^{2}}
     + \frac{\Xii\,(\alpha_j^{2}-\alpha_i^{2})\,\ell_i}{\bar R_i}
     + \Lambda \alpha_i \Xia\Xib\, \Ecal
       \left( \frac{\xi^{2}}{\Xir} + \frac{\bar\chi}{\Dth} \right)
     + \frac{\bar\kappa(\xi)\, \alpha_i \Xii \Psi(\xi)}{\Xir \bar R_i} \,.
  \label{eq:bar-phi}
\end{align}

Note, first of all, that the passage from \eqref{eq:zeta-rtheta}--\eqref{eq:zeta-phi} to
\eqref{eq:bar-rtheta}--\eqref{eq:bar-phi} is nothing but the replacement of every
symbol by its rescaled counterpart together with $2M \to 1$ and no equation changes
shape. Note also that the nine parameters $(M, a, b, \lambda; E, L_1, L_2, K, \eps)$
of the original system enter \eqref{eq:bar-rtheta}--\eqref{eq:bar-phi} only through the
eight dimensionless combinations $(\alpha, \beta, \Lambda; E, \ell_1, \ell_2, \Kcal, \eps)$,
the overall scale having been absorbed. In particular, the Myers--Perry limit of
Sec.~\ref{sec:tauphi} is now simply $\Lambda \to 0$. Finally, because the two causal
characters are carried by $\eps$ rather than by a particle mass, the rescaling
\eqref{eq:rescale1}--\eqref{eq:rescale2} needs no auxiliary mass parameter and applies
verbatim to null geodesics. The Kerr case of \cite{Cieslik:2023qdc} is different. There
the constants of motion are made dimensionless with the rest mass $m$, and the null case
requires $m$ to be traded for an arbitrary positive $\tilde m$.

%%-------------------------------------------------------------------- IV.E --
\subsection{The radial and polar polynomials}\label{sec:polynomials}

The two equations \eqref{eq:bar-rtheta} become polynomial in the squared variables
\begin{equation} \label{eq:xu}
  x := \xi^{2} \,, \qquad u := \cos^{2}\theta \,, \qquad \delta := \alpha^{2}-\beta^{2} \,,
\end{equation}
for which it is convenient to name the cubic
\begin{equation} \label{eq:Deltabar}
  \bar\Delta(x) := \Xir \bar R_1 \bar R_2 - x = \xi^{2}\,(\bar V - 1) \,,
\end{equation}
whose positive simple roots locate the horizons, and to record that
\begin{equation} \label{eq:linearinu}
  \Dth = \Xia\,u + \Xib\,(1-u) \,, \qquad
  \bar\chi = \alpha^{2} u + \beta^{2}(1-u) \,,
\end{equation}
are both \emph{linear} in $u$, the first because $\Lambda\delta = \Xia - \Xib$. Since
$\dd x/\dd\gamma = 2\xi\,\dd\xi/\dd\gamma$ and
$\dd u/\dd\gamma = -2\sin\theta\cos\theta\,\dd\theta/\dd\gamma$, the pair
\eqref{eq:bar-rtheta} is equivalent to
\begin{equation} \label{eq:PQode}
  \left( \frac{\dd x}{\dd\gamma} \right)^{\!2} = 4\,P(x) \,,
  \qquad
  \left( \frac{\dd u}{\dd\gamma} \right)^{\!2} = 4\,Q(u) \,,
\end{equation}
with $P := x\,\bar\Rcal$ and $Q := \sin^{2}\!\theta\,\cos^{2}\!\theta\;\bar\Tcal$, that is
\begin{align}
  P(x) &= \bar\Delta(x)
     \left[ \eps x - \Kcal
          + \delta \left( \frac{\Xia \ell_1^{2}}{\bar R_1}
                        - \frac{\Xib \ell_2^{2}}{\bar R_2} \right)
          + \frac{\Xia\Xib\,\Ecal^{2}\,x}{\Xir} \right]
     + \Xir \bar R_1 \bar R_2\,\Psi(x)^{2} \,,
  \label{eq:Pdef}\\[4pt]
  Q(u) &= \Dth
     \Bigl[ (\Kcal + \eps\,\bar\chi)\,u(1-u)
          - \Xia \ell_1^{2}\, u - \Xib \ell_2^{2}\,(1-u) \Bigr]
     + \Xia\Xib\,\Ecal^{2}\,\bar\chi\,u(1-u) \,.
  \label{eq:Qdef}
\end{align}
That $Q$ is a polynomial is manifest from \eqref{eq:linearinu}. That $P$ is one is not,
since $\Xir$, $\bar R_1$ and $\bar R_2$ all appear in denominators in \eqref{eq:Pdef}. Each
of the three divides out. For $\Xir$ this is the cancellation already met below
\eqref{eq:Vtilde}, between the $\Xia\Xib\Ecal^{2}x/\Xir$ term and the pole of $\Psi^{2}$.
For $\bar R_i$ it is the analogous cancellation between the $\delta$ bracket and the pole of
$\Psi$ at $x = -\alpha_i^{2}$. Consequently $\bar\Rcal = P(\xi^{2})/\xi^{2}$ carries no
singularity other than the one at the origin, in particular none at the de~Sitter radius
$\xi^{2} = 1/\Lambda$.

In fact, both $P$ and $Q$ are \emph{quartic}. Writing
\begin{equation} \label{eq:PQexpand}
  P(x) = \sum_{k=0}^{4} p_k\, x^{k} \,, \qquad Q(u) = \sum_{k=0}^{4} q_k\, u^{k} \,,
\end{equation}
their coefficients are, in the dimensionless parameters and with the co-rotating energy
$\Ecal$ of \eqref{eq:barkappa} used in place of $E$,
\begin{subequations} \label{eq:pk}
\begin{align}
  p_4 &= -\Lambda\eps \,, \label{eq:p4}\\[4pt]
  p_3 &= \Xia\Xib\,\Ecal^{2} + \Lambda\Kcal
       + \bigl[\, 1 - \Lambda(\alpha^{2}+\beta^{2}) \,\bigr]\eps \,, \label{eq:p3}\\[4pt]
  p_2 &= (\alpha^{2}+\beta^{2})\,\Xia\Xib\,\Ecal^{2}
       - \Lambda\delta \bigl( \Xia\ell_1^{2} - \Xib\ell_2^{2} \bigr)
       - \bigl[\, 1 - \Lambda(\alpha^{2}+\beta^{2}) \,\bigr]\Kcal
       + \bigl[\, \alpha^{2}+\beta^{2}-1-\Lambda\alpha^{2}\beta^{2} \,\bigr]\eps \,,
       \label{eq:p2}\\[4pt]
  p_1 &= \bigl[\, \alpha^{2}\beta^{2}\,\Xia\Xib
              + \Lambda\alpha^{2}\beta^{2} + \alpha^{2}+\beta^{2} \,\bigr]\Ecal^{2}
       - 2\Ecal\,(\alpha\ell_1 + \beta\ell_2) - 2\Lambda\alpha\beta\,\ell_1\ell_2
       \nonumber\\[2pt]
      &\quad + \bigl[\, \delta\,\Xia(2-\Xib) - \Lambda\alpha^{2} \,\bigr]\ell_1^{2}
       - \bigl[\, \delta\,\Xib(2-\Xia) + \Lambda\beta^{2} \,\bigr]\ell_2^{2}
       \nonumber\\[2pt]
      &\quad + \bigl[\, 1 - \alpha^{2}-\beta^{2} + \Lambda\alpha^{2}\beta^{2} \,\bigr]\Kcal
       + \alpha^{2}\beta^{2}\,\eps \,, \label{eq:p1}\\[4pt]
  p_0 &= \alpha^{2}\beta^{2}\bigl( \Ecal^{2} - \Kcal \bigr)
       - 2\alpha\beta\bigl( \beta\Ecal\ell_1 + \alpha\Ecal\ell_2 - \ell_1\ell_2 \bigr)
       + \beta^{2}\bigl( 1 + \delta\,\Xia \bigr)\ell_1^{2}
       + \alpha^{2}\bigl( 1 - \delta\,\Xib \bigr)\ell_2^{2} \,, \label{eq:p0}
\end{align}
\end{subequations}
and
\begin{subequations} \label{eq:qk}
\begin{align}
  q_4 &= -\Lambda\delta^{2}\eps \,, \label{eq:q4}\\[4pt]
  q_3 &= -\delta \Bigl[\, \Xia\Xib\,\Ecal^{2} + \Lambda\Kcal
       + \bigl( 1 + 3\Lambda\beta^{2} - \Lambda\alpha^{2} \bigr)\eps \,\Bigr] \,,
       \label{eq:q3}\\[4pt]
  q_2 &= (\alpha^{2}-2\beta^{2})\,\Xia\Xib\,\Ecal^{2}
       - \Lambda\delta \bigl( \Xia\ell_1^{2} - \Xib\ell_2^{2} \bigr)
       - \bigl( 2\Xib - \Xia \bigr)\Kcal
       \nonumber\\[2pt]
      &\quad + \bigl[\, \alpha^{2}-2\beta^{2}
                      + \Lambda\beta^{2}(2\alpha^{2}-3\beta^{2}) \,\bigr]\eps \,,
       \label{eq:q2}\\[4pt]
  q_1 &= \Xib \Bigl[\, \beta^{2}\Xia\,\Ecal^{2} + \Kcal + \beta^{2}\eps
       - \Xia\ell_1^{2} + \bigl( 2\Xib - \Xia \bigr)\ell_2^{2} \,\Bigr] \,,
       \label{eq:q1}\\[4pt]
  q_0 &= -\Xib^{2}\ell_2^{2} \,. \label{eq:q0}
\end{align}
\end{subequations}
The two endpoints of $Q$ are the turning points on the axes: $q_0 = Q(0) = -\Xib^{2}\ell_2^{2}$
and $Q(1) = -\Xia^{2}\ell_1^{2}$. Equivalently,
\begin{equation} \label{eq:P0compact}
  p_0 = -\alpha^{2}\beta^{2}\Kcal
      + \delta \bigl( \Xia \beta^{2}\ell_1^{2} - \Xib \alpha^{2}\ell_2^{2} \bigr)
      + \bigl( \alpha\beta\,\Ecal - \beta\ell_1 - \alpha\ell_2 \bigr)^{2} \,.
\end{equation}

Both leading coefficients are proportional to $\Lambda\eps$, so the quartic behavior is a
joint effect of the cosmological constant and a nonzero rest mass. The degenerations are
collected in Table~\ref{tab:degrees}.
\begin{table}[htb]
\caption{Degrees of the radial and polar polynomials
\eqref{eq:Pdef}--\eqref{eq:Qdef} in the various limits, and the resulting class of
functions. ``Elementary'' means that the corresponding quadrature is solved by
trigonometric functions.}
\label{tab:degrees}
\begin{ruledtabular}
\begin{tabular}{lccl}
 & $\deg P$ & $\deg Q$ & solved by \\
\hline
 generic: \ $\Lambda \neq 0$, \ $\eps = -1$, \ $\alpha \neq \beta$
   & $4$ & $4$ & elliptic / elliptic \\
 null: \ $\eps = 0$
   & $3$ & $3$ & elliptic / elliptic \\
 Myers--Perry: \ $\Lambda = 0$
   & $3$ & $3$ & elliptic / elliptic \\
 equal rotations: \ $\alpha = \beta$
   & $4$ & $2$ & elliptic / elementary \\
 $\Lambda = 0$ and $\alpha = \beta$
   & $3$ & $2$ & elliptic / elementary \\
\end{tabular}
\end{ruledtabular}
\end{table}

It is worth setting Table~\ref{tab:degrees} against the four-dimensional situation. In Kerr
the radial polynomial is quartic and the motion is elliptic. Switching on a cosmological
constant raises it to degree six for timelike geodesics. Kerr--(A)dS therefore requires
hyperelliptic Kleinian $\sigma$ functions restricted to the $\theta$-divisor, and the
quartic, hence elliptic, case survives only for null geodesics
\cite{Hackmann:2010zz,Hackmann:thesis}. In five
dimensions the same step is far milder: because the natural radial variable is $x = r^{2}$,
the Myers--Perry polynomial is a cubic \cite{Diemer:2014lba} and the cosmological constant
raises it only to a quartic. Both sectors therefore remain elliptic in five dimensions, for
timelike as well as null geodesics, where their four-dimensional counterparts do not. The
comparison is collected in Table~\ref{tab:degrees4d}.
\begin{table}[htb]
\caption{The four-dimensional counterpart of Table~\ref{tab:degrees}, for the Kerr and
Kerr--(A)dS black holes, read off from Eqs.~(4.1.7)--(4.1.8) of \cite{Hackmann:thesis} in
the convention of Table~\ref{tab:degrees}: the entries are the degrees of the polynomials
standing on the right-hand sides of $(\dd r/\dd\gamma)^{2}$ and $(\dd\nu/\dd\gamma)^{2}$,
with $\nu = \cos^{2}\theta$ and $\gamma$ the Mino parameter. The polar degree usually
quoted, as in \cite{Hackmann:2010zz}, is that of $\Theta_\nu = (1-\nu)\Theta$ and is one
lower, the equation of motion being $(\dd\nu/\dd\gamma)^{2} = 4\nu\,\Theta_\nu$.}
\label{tab:degrees4d}
\begin{ruledtabular}
\begin{tabular}{lccl}
 & radial & polar & solved by \\
\hline
 Kerr, \ timelike             & $4$ & $3$ & elliptic / elliptic \\
 Kerr, \ null                 & $4$ & $3$ & elliptic / elliptic \\
 Kerr--(A)dS, \ timelike      & $6$ & $4$ & hyperelliptic / elliptic \\
 Kerr--(A)dS, \ null          & $4$ & $3$ & elliptic / elliptic \\
\end{tabular}
\end{ruledtabular}
\end{table}
The single entry of degree six is what forces the Kleinian $\sigma$ functions of
\cite{Hackmann:2010zz}. Every other entry in Tables~\ref{tab:degrees} and
\ref{tab:degrees4d}, in four dimensions and in five, is of degree three or four, and so
falls within the reach of the construction of Sec.~\ref{sec:bw}.

%%-------------------------------------------------------------------- IV.F --
\subsection{Relation to Grunau, Neumann and Reimers}\label{sec:earlier}

The geodesics of the five-dimensional Kerr--AdS black hole have been studied in detail by
Grunau, Neumann and Reimers \cite{Grunau:2017uwv}. Working with two independent rotation
parameters and a negative cosmological constant, they pass to $x = r^{2}$ and
$\nu = \cos^{2}\theta$, introduce the same Mino-type parameter as in Sec.~\ref{sec:repara},
obtain quartic polynomials in both sectors, and integrate them in terms of the Weierstrass
$\wp$, $\sigma$ and $\zeta$ functions. They then classify the orbits. The turning points are
the zeros of the radial quartic, and the boundaries in parameter space are its double zeros
together with $P(-\alpha^{2}) = 0$. From the resulting parametric diagrams, and from an
effective potential $V_\pm$ obtained by solving $P = 0$ for the energy, they enumerate the
possible orbit types for timelike, null and spacelike geodesics.

Our equations agree with theirs exactly. The dictionary is
\begin{equation} \label{eq:GNRdict}
  g^{2} = -\Lambda \,, \qquad
  \delta_{\rm GNR} = -\eps \,, \qquad
  E_{\rm GNR} = \Ecal \,, \qquad
  K_{\rm GNR} = \Kcal + \Lambda \alpha^{2}\beta^{2}\Ecal^{2} \,,
\end{equation}
under which Eq.~(15) of \cite{Grunau:2017uwv} is $4P(x)$ and its Eq.~(19) is $Q(u)$,
identically in $x$ and $u$ respectively. The third identification has a geometric origin
worth recording. The metric of \cite{Grunau:2017uwv} is written in a frame that does not
rotate at infinity, whereas \eqref{eq:ds2} is not, and the two are related by
\begin{equation} \label{eq:GNRangles}
  t_{\rm GNR} = \tau \,, \qquad
  \phi_{\rm GNR} = \ph_1 - \lambda a\,\tau \,, \qquad
  \psi_{\rm GNR} = \ph_2 - \lambda b\,\tau \,,
\end{equation}
as one verifies on \eqref{eq:ds2} directly. The momentum conjugate to $t_{\rm GNR}$ at
fixed $\phi_{\rm GNR}$ and $\psi_{\rm GNR}$ is then $E - \lambda a_i L_i = \Ecal$, while
$L_1$ and $L_2$ are unchanged. We return to \eqref{eq:GNRangles} in
Sec.~\ref{sec:azimuthalsol}, where it accounts for a difference between the two sets of
azimuthal quadratures. Two further points deserve emphasis. First, the last two
identifications in \eqref{eq:GNRdict}, i.e., that their energy is the co-rotating $\Ecal$ and
that their separation constant is a third normalization of the Carter constant, are
forced independently by the radial and by the polar comparison, which is a nontrivial
consistency check on both calculations. Reference \cite{Grunau:2017uwv} does not cite
\cite{Kunduri:2005fq}, so its derivation is independent. Its agreement with the corrected
$\widetilde{V}$ of \eqref{eq:Vtilde} is therefore a third confirmation of the correction of
Sec.~\ref{sec:compare}, alongside \cite{Davis:2005ys} and \cite{Wu:2008bc}. Second, of the
three normalizations of the separation constant now in circulation, $\Kkl$ is the only one
that fails to have a limit as $\lambda \to 0$. Ours and that of \cite{Grunau:2017uwv} differ
from each other at $O(\lambda)$, and both reduce to the constant of \cite{Frolov:2003en}.

What we do differently, from Sec.~\ref{sec:bw} on, is the integration itself. The treatment
of \cite{Grunau:2017uwv} proceeds case by case, the form of the solution depending on the
number and the position of the real zeros of $P$ and $Q$. For the four-dimensional Kerr
spacetime, Cie\'slik, Hackmann and Mach \cite{Cieslik:2023qdc} have shown that a
little-known result of Biermann and Weierstrass removes this case splitting: a single set of
formulas, explicitly real for every orbit type, parametrized by the constants of motion and
the initial coordinates. Equations \eqref{eq:PQode} are of exactly the form that result
requires, in both sectors at once, and the rest of this paper is devoted to carrying the
construction of \cite{Cieslik:2023qdc} over to them. As far as we know, the closest existing
application to what we do here is that of Faraji and Hackmann \cite{Faraji:2025xkl}, who
use dimensionless variables and a Mino-type parametrization to bring both the radial and
the polar sector of the accelerated Schwarzschild black hole into Biermann--Weierstrass
form. Their setting is four-dimensional and static, with vanishing cosmological constant,
and restricted to null geodesics. The five-dimensional rotating case with $\Lambda \neq 0$
and $\eps = -1$, in which both polynomials are genuinely quartic, has not been treated
in this way. Everything recorded in Secs.~\ref{sec:hj} and \ref{sec:firstorder} is the corrected
and streamlined form on which that construction is built.

%%%%%%%%%%%%%%%%%%%%%%%%%%%%%%%%%%%%%%%%%%%%%%%%%%%%%%%%%%%%%%%%%%%%%%%%%%%%%%
%%  V.  THE BIERMANN-WEIERSTRASS SOLUTION
%%%%%%%%%%%%%%%%%%%%%%%%%%%%%%%%%%%%%%%%%%%%%%%%%%%%%%%%%%%%%%%%%%%%%%%%%%%%%%
\section{Solution in terms of Weierstrass functions}\label{sec:bw}

This is the body of the paper. The system \eqref{eq:PQode} is integrated in closed form:
first the radial and polar coordinates, by a theorem of Biermann and Weierstrass that takes
an arbitrary point of the orbit as its initial datum, and then the time and the two
azimuthal angles, by quadratures that turn out to involve elliptic integrals of the third
kind and nothing else. The affine parameter that the Mino parameter displaced is recovered
next, and the last subsection takes the degenerate cases of Table~\ref{tab:degrees}.

%%--------------------------------------------------------------------- V.A --
\subsection{The Biermann--Weierstrass theorem}\label{sec:bwthm}

We first state the theorem in the form we need, say why it is the right tool here, and
record the invariants of our two quartics. Both of the equations \eqref{eq:PQode} are of
the form
\begin{equation} \label{eq:bwform}
  \left( \frac{\dd y}{\dd z} \right)^{\!2} = f(y) \,, \qquad y(0) = y_0 \,,
\end{equation}
with $f$ a quartic: for the radial sector $y = x$ and $f = 4P$, and for the polar sector
$y = u$ and $f = 4Q$. In both cases the independent variable is the same, namely
\begin{equation} \label{eq:zdef}
  z = \gamma - \gamma_0 \,,
\end{equation}
because $\dd\gamma = \pm\,\dd y/\sqrt{f(y)}$ in either sector. Here $y_0$ is the value of
$y$ at $\gamma = \gamma_0$, and $\gamma_0$ may be taken to be the same for the two sectors. The
radial and polar motions are therefore two Weierstrass functions of one and the same
argument, differing only in their invariants.

Write the quartic in the classical normalization
\begin{equation} \label{eq:quarticnorm}
  f(y) = a_0\,y^{4} + 4a_1\,y^{3} + 6a_2\,y^{2} + 4a_3\,y + a_4 \,,
\end{equation}
and let
\begin{equation} \label{eq:g2g3}
  g_2 = a_0a_4 - 4a_1a_3 + 3a_2^{2} \,, \qquad
  g_3 = a_0a_2a_4 + 2a_1a_2a_3 - a_2^{3} - a_0a_3^{2} - a_1^{2}a_4
\end{equation}
be its two invariants. The theorem of Biermann and Weierstrass
\cite{Biermann:1865,WhittakerWatson,Greenhill:1892}, in the form used in
\cite{Cieslik:2022uki,Cieslik:2023qdc}, then states that the solution of \eqref{eq:bwform}
is
\begin{equation} \label{eq:bw}
  y(z) = y_0
    + \frac{ -\,s\,\wp'(z)
           + \dfrac{1}{2}f'(y_0)\left[\, \wp(z) - \dfrac{1}{24}f''(y_0) \,\right]
           + \dfrac{1}{24}\, f(y_0)\, f'''(y_0) }
           { 2\left[\, \wp(z) - \dfrac{1}{24}f''(y_0) \,\right]^{2}
           - \dfrac{1}{48}\, f(y_0)\, f^{(4)}(y_0) } \,,
\end{equation}
where $\wp(z) := \wp(z; g_2, g_3)$ is the Weierstrass function with the invariants
\eqref{eq:g2g3} and $s$ is either square root of $f(y_0)$. If the initial point is chosen at a turning point, so
that $f(y_0) = 0$, \eqref{eq:bw} collapses to
\begin{equation} \label{eq:bwturn}
  y(z) = y_0 + \frac{f'(y_0)}{4\wp(z) - \dfrac{1}{6}f''(y_0)} \,.
\end{equation}

Three features of \eqref{eq:bw} are what make it the appropriate tool here. First, $y_0$ is
an arbitrary point of the orbit and is \emph{not} required to be a zero of $f$: the initial
position enters the solution explicitly, on the same footing as the constants of motion,
and no auxiliary construction is needed to start the trajectory at a generic point.
Second, and most importantly, \eqref{eq:bw} is manifestly real. The coefficients of $P$ and
$Q$ listed in \eqref{eq:pk}--\eqref{eq:qk} are real, hence so are $g_2$ and $g_3$. The
Laurent coefficients of $\wp$ are polynomials in $g_2$ and $g_3$ with rational coefficients,
so that $\overline{\wp(z;g_2,g_3)} = \wp(\bar z;g_2,g_3)$ and both $\wp$ and $\wp'$ are real
for real $z$. Moreover $s$ is real, because $f(y_0) \geq 0$ at any point the orbit actually
visits. Equation \eqref{eq:bw} is therefore real for real $z$ whatever the number of real
zeros of $f$, and the solution need not be split into cases according to the root structure
of the polynomial, which is precisely the case splitting that organizes the treatment of
\cite{Grunau:2017uwv}. Third, neither \eqref{eq:g2g3} nor \eqref{eq:bw} divides by $a_0$,
so both continue to hold when the quartic degenerates to a cubic, and indeed to a quadratic.
A single formula therefore covers every row of Table \ref{tab:degrees}. The elementary
entries appear as the limit in which the period lattice itself degenerates, as we shall
see in Sec.~\ref{sec:polarsol}.

It remains to record the invariants of our two quartics. Matching
$4P(x) = \sum_k 4p_k x^{k}$ against \eqref{eq:quarticnorm} gives
\begin{equation} \label{eq:akfrompk}
  a_0 = 4p_4 \,, \qquad a_1 = p_3 \,, \qquad a_2 = \frac{2}{3}\,p_2 \,, \qquad
  a_3 = p_1 \,, \qquad a_4 = 4p_0 \,,
\end{equation}
whence, writing $g_2^{(r)}$ and $g_3^{(r)}$ for the radial invariants,
\begin{equation} \label{eq:ginvP}
\begin{split}
  g_2^{(r)} &= \frac{4}{3}\,\mathcal{I}[P] \,, \qquad\qquad
  g_3^{(r)} = \frac{4}{27}\,\mathcal{J}[P] \,, \\[4pt]
  \mathcal{I}[P] &:= p_2^{2} - 3p_1p_3 + 12p_0p_4 \,, \\[2pt]
  \mathcal{J}[P] &:= 72p_0p_2p_4 + 9p_1p_2p_3 - 27p_1^{2}p_4 - 27p_0p_3^{2} - 2p_2^{3} \,,
\end{split}
\end{equation}
and $g_2^{(\theta)}, g_3^{(\theta)}$ are given by the same expressions with $q_k$ in place of
$p_k$. The two brackets $\mathcal{I}$ and $\mathcal{J}$ are the classical invariants of a
quartic, and the discriminant of $P$ is $\bigl(4\mathcal{I}^{3}-\mathcal{J}^{2}\bigr)/27$.
Consequently
$\bigl( g_2^{(r)} \bigr)^{3} - 27 \bigl( g_3^{(r)} \bigr)^{2}
 = 16\bigl(4\mathcal{I}^{3}-\mathcal{J}^{2}\bigr)/27$,
and likewise for the polar sector. The sign of the modular discriminant, which decides
whether the period lattice of $\wp$ is rectangular or rhombic, is thus simply the sign of
the discriminant of the corresponding orbital polynomial: positive when it has four real
roots or none, negative when it has exactly two.

%%--------------------------------------------------------------------- V.B --
\subsection{The radial solution}\label{sec:radialsol}

For the radial sector take $y = x$ and $f = 4P$ in \eqref{eq:bwform}--\eqref{eq:bw}. Since
$P$ is a quartic its two highest derivatives are almost trivial,
\begin{equation} \label{eq:Pderivs}
  P'''(x_0) = 6\,\bigl( p_3 + 4p_4 x_0 \bigr) \,, \qquad
  P^{(4)}(x_0) = 24\,p_4 \,,
\end{equation}
so only the two combinations $p_3 + 4p_4x_0$ and $p_4$ survive from the last term of the
numerator and from the denominator of \eqref{eq:bw}. Abbreviating the shifted Weierstrass
function by
\begin{equation} \label{eq:wpstar}
  \wp_{\!*}^{(r)}(z) := \wp\bigl( z;\, g_2^{(r)},\, g_3^{(r)} \bigr)
                - \frac{1}{6}\,P''(x_0) \,,
  \qquad z = \gamma - \gamma_0 \,,
\end{equation}
with the invariants \eqref{eq:ginvP}, and canceling a common factor of two between
numerator and denominator, \eqref{eq:bw} becomes
\begin{equation} \label{eq:xsol}
  x(\gamma) = x_0
  + \frac{ P'(x_0)\,\wp_{\!*}^{(r)}(z)
         \;-\; s_r \sqrt{P(x_0)}\;\wp'\bigl( z;\, g_2^{(r)},\, g_3^{(r)} \bigr)
         \;+\; 2\,P(x_0)\,\bigl( p_3 + 4p_4 x_0 \bigr) }
         { \bigl[ \wp_{\!*}^{(r)}(z) \bigr]^{2} \;-\; 4\,p_4\,P(x_0) } \,,
\end{equation}
in which
\begin{equation} \label{eq:Pat0}
  P(x_0) = \sum_{k=0}^{4} p_k\,x_0^{k} \,, \qquad
  P'(x_0) = p_1 + 2p_2x_0 + 3p_3x_0^{2} + 4p_4x_0^{3} \,, \qquad
  P''(x_0) = 2p_2 + 6p_3x_0 + 12p_4x_0^{2} \,,
\end{equation}
and $s_r = \pm 1$. The radial coordinate itself is recovered from $x = \xi^{2}$ and
$\xi = r/r_0$, so that $r(\gamma) = r_0\sqrt{x(\gamma)}$. The motion is confined to
$x \geq 0$.

Equation \eqref{eq:xsol} is the complete radial solution. It is parametrized by the eight
dimensionless constants $(\alpha, \beta, \Lambda; E, \ell_1, \ell_2, \Kcal, \eps)$, which
enter only through the coefficients $p_k$ of \eqref{eq:pk} and through the invariants
\eqref{eq:ginvP}, and by the three initial data $\gamma_0$, $x_0$ and $s_r$. The last
of these fixes the initial direction of the radial motion: differentiating
\eqref{eq:xsol} at $z = 0$, where $\wp(z) = z^{-2} + O(z^{2})$ and
$\wp'(z) = -2z^{-3} + O(z)$, gives
\begin{equation} \label{eq:xinit}
  x(\gamma_0) = x_0 \,, \qquad
  \left. \frac{\dd x}{\dd \gamma} \right|_{\gamma_0} = 2\,s_r \sqrt{P(x_0)} \,,
\end{equation}
so that $s_r$ is the sign of $\dd\xi/\dd\gamma$ at $\gamma = \gamma_0$. Since $\wp$ is
even and $\wp'$ is odd, reversing $s_r$ is the same as reversing $z$.

If the starting point is taken at a radial turning point, $P(x_0) = 0$, the square root and
the last term of the numerator both vanish and \eqref{eq:xsol} collapses to
\begin{equation} \label{eq:xsolturn}
  x(\gamma) = x_0 + \frac{P'(x_0)}{\wp_{\!*}^{(r)}(z)} \,,
\end{equation}
which is \eqref{eq:bwturn} written for $f = 4P$. This is the form most convenient for
bounded radial motion, which oscillates between two consecutive simple zeros of $P$: the
smaller of the two is the \emph{periastron}, the point of the orbit lying closest to the
black hole, and the larger is the \emph{apastron}, the point lying farthest from it. The
radial velocity vanishes at both, so for such an orbit $\gamma_0$ may always be placed at
one of them and \eqref{eq:xsolturn} used in place of \eqref{eq:xsol}. The general form
\eqref{eq:xsol} is what is needed when the orbit is to be started at a prescribed interior
point, or when it has no turning point at all.

Two properties follow at once. First, $x(\gamma)$ is a rational function of $\wp$ and
$\wp'$ and is therefore itself an elliptic function of $\gamma$, with the period lattice of
$\wp\bigl(z; g_2^{(r)}, g_3^{(r)}\bigr)$. When the radial motion is bounded between two
simple real zeros of $P$ it is consequently periodic in the Mino parameter, the radial
period being the real period $2\omega$ of that lattice. Second, nothing in \eqref{eq:xsol}
degenerates when $p_4 = 0$: the denominator reduces to
$\bigl[\wp_{\!*}^{(r)}\bigr]^{2}$ and the numerator to
$P'(x_0)\wp_{\!*}^{(r)} - s_r\sqrt{P(x_0)}\,\wp' + 2P(x_0)p_3$. By \eqref{eq:p4} this is the
case for null geodesics and in the Myers--Perry limit, so the single expression
\eqref{eq:xsol} serves every row of Table \ref{tab:degrees}.

It is worth setting \eqref{eq:xsol} against the radial solution of
\cite{Grunau:2017uwv}. There the quartic is brought to Weierstrass form by the classical
route: the substitution $x = x_R + 1/y$ reduces it to a cubic in $y$ with coefficients
$b_3 = X'(x_R)$, $b_2 = X''(x_R)/2$, $b_1 = X'''(x_R)/6$ and $b_0 = X^{(4)}(x_R)/24$, where
$X = 4P$, and an affine map of $y$ then yields their Eq.~(39),
\begin{equation} \label{eq:gnrsol}
  x(\gamma) = x_R
   + \frac{b_3}{4\wp\bigl(\gamma - \gamma_{\rm in}'; g_2^{x}, g_3^{x}\bigr) - b_2/3} \,,
\end{equation}
with $g_2^{x} = b_2^{2}/12 - b_1b_3/4$ and
$g_3^{x} = b_1b_2b_3/48 - b_0b_3^{2}/16 - b_2^{3}/216$. Their $\gamma_{\rm in}'$ is the
integration constant that fixes the origin of the Mino parameter, our $\gamma_0$. The
substitution requires $x_R$ to be a \emph{zero} of $X$. Given that, \eqref{eq:gnrsol} is identical to our
\eqref{eq:xsolturn}, and the invariants $g_2^{x}, g_3^{x}$, assembled there from the
expansion of $X$ about $x_R$, coincide with \eqref{eq:ginvP}, which is built from the
$p_k$ alone and makes no reference to any root. This is as it should be: $\mathcal{I}$ and
$\mathcal{J}$ are unaltered by a translation of the argument, so the choice of zero is
immaterial.

\begin{figure}[tbp]
  \centering
  \includegraphics[width=0.86\columnwidth]{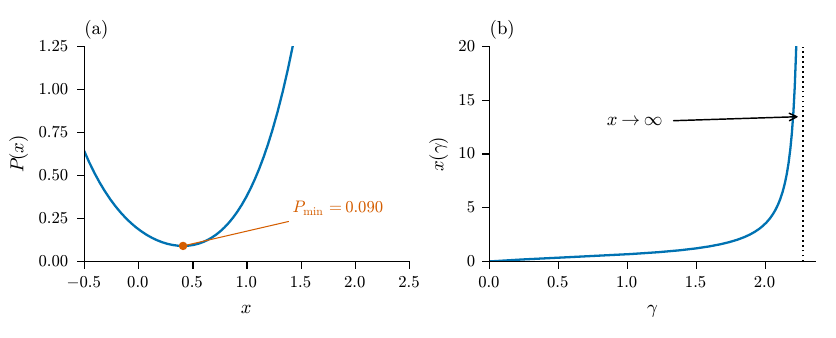}
  \caption{A de~Sitter case in which the radial quartic has no real zero:
    $\alpha = 1/4$, $\beta = 3/10$, $\Lambda = 3/10$, $\eps = -1$, $E = 9/10$,
    $\ell_1 = 4/5$, $\ell_2 = 1$ and $\Kcal = 3/10$. (a) $P(x)$ is positive throughout,
    its real minimum being $0.0903$ at $x = 0.4094$. Its four zeros are
    $0.4408 \pm 0.3931\,i$ and $0.0188 \pm 1.3353\,i$. (b) the branch of $x(\gamma)$ that
    leaves the origin, computed from \eqref{eq:xsol} with $x_0 = 1$ and $s_r = +1$, the
    origin of $\gamma$ being placed where $x$ vanishes. The orbit has no radial turning
    point: $x$ rises without bound as $\gamma$ approaches the finite value marked by the
    dotted line. The substitution leading to \eqref{eq:gnrsol} therefore has no real
    starting point here. Drawn with a Python script written with the assistance
    disclosed in the Acknowledgments; the values quoted were verified independently.}
  \label{fig:nots}
\end{figure}

Equation \eqref{eq:xsol} is therefore the extension of \eqref{eq:gnrsol} to an arbitrary
starting point, and the extension is not merely a convenience. If $P$ has no real zero the
substitution $x = x_R + 1/y$ has no real starting point at all, and this case does arise.
For $\Lambda < 0$ one has $p_4 = -\Lambda\eps < 0$ for timelike geodesics by
\eqref{eq:p4}, so that $P \to -\infty$ at both ends and a real zero necessarily exists
whenever any radial motion is possible. The same is true of the cubic left when
$\eps = 0$. But for $\Lambda > 0$ one has $p_4 > 0$, and $P$ may then be positive for every
$x$, the orbit having no radial turning point whatever.\footnote{For instance
$\alpha = 1/4$, $\beta = 3/10$, $\Lambda = 3/10$, $E = 9/10$, $\ell_1 = 4/5$,
$\ell_2 = 1$ and $\Kcal = 3/10$ give a $P$ whose four zeros are all complex, with
$P > 0$ throughout. This is the case drawn in Fig.~\ref{fig:nots}.}
Reference \cite{Grunau:2017uwv} treats $\Lambda < 0$ only and does
not meet this situation. Equation \eqref{eq:xsol} covers it without modification, and does
so with the same formula that serves every other case.

%%--------------------------------------------------------------------- V.C --
\subsection{The polar solution}\label{sec:polarsol}

The polar sector is treated in exactly the same way, with $y = u = \cos^{2}\theta$ and
$f = 4Q$. Since $Q$ is again a quartic,
\begin{equation} \label{eq:Qderivs}
  Q'''(u_0) = 6\,\bigl( q_3 + 4q_4 u_0 \bigr) \,, \qquad Q^{(4)}(u_0) = 24\,q_4 \,,
\end{equation}
and writing, in parallel with \eqref{eq:wpstar},
\begin{equation} \label{eq:wpstartheta}
  \wp_{\!*}^{(\theta)}(z) := \wp\bigl( z;\, g_2^{(\theta)},\, g_3^{(\theta)} \bigr)
                           - \frac{1}{6}\,Q''(u_0) \,,
  \qquad z = \gamma - \gamma_0 \,,
\end{equation}
Eq.~\eqref{eq:bw} gives
\begin{equation} \label{eq:usol}
  u(\gamma) = u_0
  + \frac{ Q'(u_0)\,\wp_{\!*}^{(\theta)}(z)
         \;-\; s_\theta \sqrt{Q(u_0)}\;
               \wp'\bigl( z;\, g_2^{(\theta)},\, g_3^{(\theta)} \bigr)
         \;+\; 2\,Q(u_0)\,\bigl( q_3 + 4q_4 u_0 \bigr) }
         { \bigl[ \wp_{\!*}^{(\theta)}(z) \bigr]^{2} \;-\; 4\,q_4\,Q(u_0) } \,,
\end{equation}
with $s_\theta = \pm 1$ and $Q(u_0)$, $Q'(u_0)$, $Q''(u_0)$ obtained from the $q_k$ of
\eqref{eq:qk} exactly as in \eqref{eq:Pat0}. Because the polar coordinate is restricted to
$\theta \in [0,\pi/2]$, the cosine does not change sign and the inversion is unambiguous,
\begin{equation} \label{eq:thetasol}
  \theta(\gamma) = \arccos \sqrt{u(\gamma)} \,.
\end{equation}
The initial data enter as before: $u(\gamma_0) = u_0$ and
$\dd u/\dd\gamma|_{\gamma_0} = 2s_\theta\sqrt{Q(u_0)}$. Here, however,
$\dd u/\dd\gamma = -\sin 2\theta \; \dd\theta/\dd\gamma$, so that on $0 < \theta < \pi/2$ the
sign $s_\theta$ is \emph{minus} that of $\dd\theta/\dd\gamma$ at $\gamma_0$: the orbit
moves toward the axis $\theta = 0$ when $s_\theta = +1$.

The two ends of the polar range are governed by the endpoint values recorded in
\eqref{eq:qk},
\begin{equation} \label{eq:Qends}
  Q(0) = -\,\Xib^{2}\ell_2^{2} \,, \qquad Q(1) = -\,\Xia^{2}\ell_1^{2} \,,
\end{equation}
both of which are negative unless the corresponding angular momentum vanishes. An orbit
therefore cannot reach $\theta = \pi/2$, where the second rotation plane degenerates,
unless $\ell_2 = 0$, nor $\theta = 0$, where the first degenerates, unless $\ell_1 = 0$:
each plane presents a centrifugal barrier at its own axis. This has a useful consequence.
Since $Q$ is negative at both ends of $[0,1]$ and must be nonnegative somewhere on the
orbit, it has at least two zeros in the open interval, so the polar motion always
oscillates between two turning points $u_{\min}$ and $u_{\max}$ with
$0 \le u_{\min} < u_{\max} \le 1$. Unlike the radial motion of Sec.~\ref{sec:radialsol},
the polar motion therefore always possesses a real turning point, and the alternative form
\begin{equation} \label{eq:usolturn}
  u(\gamma) = u_0 + \frac{Q'(u_0)}{\wp_{\!*}^{(\theta)}(z)} \,,
  \qquad Q(u_0) = 0 \,,
\end{equation}
is always available. Equation \eqref{eq:usol} remains the form to use when the orbit is to
be started at a prescribed inclination $\theta_0$ rather than at a turning point.

\begin{figure}[tbp]
  \centering
  \includegraphics[width=0.805\columnwidth]{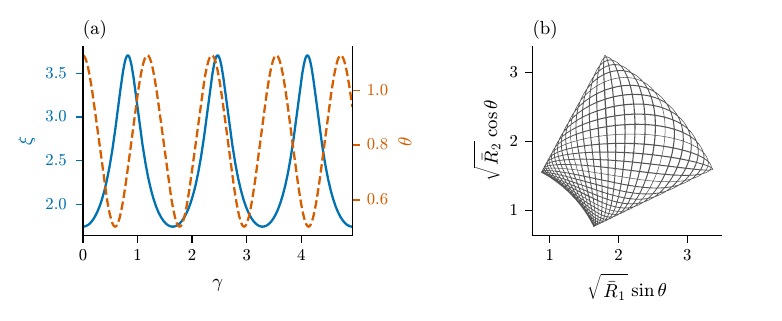}
  \caption{A bound timelike orbit of the anti-de~Sitter case, from \eqref{eq:xsolturn} and
    \eqref{eq:usolturn}, for $\alpha = 1/2$, $\beta = 3/10$, $\Lambda = -1/10$,
    $\eps = -1$, $E = 9/5$, $\ell_1 = 6/5$, $\ell_2 = 1$ and $\Kcal = 34/5$.
    (a) $\xi(\gamma)$ and $\theta(\gamma)$ over three radial periods. The radial motion
    oscillates between $\xi = 1.745$ and $\xi = 3.704$ and the polar motion between
    $\theta = 0.503$ and $\theta = 1.127$. (b) the same orbit followed for twenty-six
    radial periods, in the plane of the two rotation radii. The two periods,
    $2\omega_r = 1.6449$ and $2\omega_\theta = 1.1810$, stand in an irrational ratio, so
    the orbit does not close and fills the region bounded by the four turning values. Drawn with a Python script written with the assistance
    disclosed in the Acknowledgments; the values quoted were verified independently.}
  \label{fig:orbit}
\end{figure}

As in the radial sector, $u(\gamma)$ is an elliptic function of the Mino parameter, so the
polar motion is periodic with the real period $2\omega_\theta$ of
$\wp\bigl(z; g_2^{(\theta)}, g_3^{(\theta)}\bigr)$. The radial and polar periods are
governed by different invariants, and their ratio is in general irrational, so that a
generic orbit does not close.\footnote{For the parameters $\alpha = 1/5$, $\beta = 1/2$,
$\Lambda = 1/25$, $E = 9/10$, $\ell_1 = 7/10$, $\ell_2 = 3/10$ and $\Kcal = 7/5$, for
example, one finds $2\omega_r = 4.1895\ldots$ and $2\omega_\theta = 2.6934\ldots$, with
$\theta$ oscillating between $0.686$ and $1.227$ radians, well inside $(0,\pi/2)$.
Figure \ref{fig:orbit} shows a bound orbit of this kind.}

The comparison with \cite{Grunau:2017uwv} runs as in Sec.~\ref{sec:radialsol}. The same
classical substitution $u = u_R + 1/y$ with $u_R$ a zero of the polar quartic, followed by
an affine map, produces their Eq.~(44), which for $\Theta_\nu = 4Q$ is exactly
\eqref{eq:usolturn}. Their invariants $g_2^{\nu}, g_3^{\nu}$ again coincide with
\eqref{eq:ginvP} formed from the $q_k$. They also record the exclusion of the axes, in the
form of the statement that a geodesic cannot reach $\theta = 0$ while $\ell_1 \neq 0$ nor
$\theta = \pi/2$ while $\ell_2 \neq 0$, which here is read off from \eqref{eq:Qends}. The
difference from the radial case is that in the polar sector a real zero of the quartic is
guaranteed, so the passage from \eqref{eq:usolturn} to \eqref{eq:usol} buys convenience,
i.e. the freedom to impose initial data at an arbitrary inclination, rather than extending
the reach of the solution.

Finally, when the two rotation parameters coincide the polar quartic degenerates. Setting
$\alpha = \beta$ in \eqref{eq:qk} annihilates $q_4$ and $q_3$ together, leaving a quadratic,
and a short calculation gives
$\bigl(g_2^{(\theta)}\bigr)^{3} - 27\bigl(g_3^{(\theta)}\bigr)^{2} = 0$ identically. The
period lattice degenerates, $\wp$ reduces to an elementary function, and with it
\eqref{eq:usol}. The last two rows of Table \ref{tab:degrees} are thus not a separate
construction but the degenerate-lattice limit of the same formula.

%%--------------------------------------------------------------------- V.D --
\subsection{The time coordinate}\label{sec:timesol}

With $x(\gamma)$ and $u(\gamma)$ in hand the three cyclic coordinates follow by quadrature:
the right-hand sides of \eqref{eq:bar-tau} and \eqref{eq:bar-phi} depend on the orbit only
through $x$ and $u$, so that $T$ and $\ph_i$ are integrals of known functions of the Mino
parameter. This is the second dividend of the reparametrization of
Sec.~\ref{sec:repara}, the first having been the decoupling of the radial and polar
motions.

It is convenient to name the numerator that $\bar\kappa\,\Psi$ carries,
\begin{equation} \label{eq:Ndef}
  N(x) := \Xir \bar R_1 \bar R_2\,\Psi(x)
        = \Ecal\,\bar R_1 \bar R_2
          - \Xir \bigl( \alpha\ell_1 \bar R_2 + \beta\ell_2 \bar R_1 \bigr) \,,
\end{equation}
a quadratic in $x$, so that $\bar\kappa\,\Psi = N/\bar\Delta$ by \eqref{eq:barkappa} and
\eqref{eq:Deltabar}. Equation \eqref{eq:bar-tau} then separates into a radial and a polar
piece,
\begin{equation} \label{eq:dTsplit}
  \frac{\dd T}{\dd\gamma} = \frac{\mathcal{N}_T(x)}{\bar\Delta(x)}
     \;+\; \Xia\Xib\,\Ecal\,\frac{\bar\chi(u)}{\Dth(u)} \,,
  \qquad
  \mathcal{N}_T := \frac{\Xia\Xib\,\Ecal\,x\,\bar\Delta + N}{\Xir} \,.
\end{equation}

The division by $\Xir$ is exact, by the mechanism that already made $P$ a polynomial in
Sec.~\ref{sec:polynomials}: at $x = 1/\Lambda$ one has $\bar R_i = \Xii/\Lambda$ and hence
$\bar R_1 \bar R_2 = \Xia\Xib\,x^{2}$, which is precisely the combination the numerator
presents there. The sharp statement is the identity
\begin{equation} \label{eq:divXir}
  \bar R_1 \bar R_2 - \Xia\Xib\,x^{2}
   = \Xir\bigl[\, (\alpha^{2}+\beta^{2})\,x
                + \alpha^{2}\beta^{2}\,(1+\Lambda x) \,\bigr] \,,
\end{equation}
which gives
\begin{equation} \label{eq:NTexplicit}
  \mathcal{N}_T(x)
   = \Ecal\bigl[\, \Xia\Xib\,x\,\bar R_1 \bar R_2 + (\alpha^{2}+\beta^{2})\,x
                 + \alpha^{2}\beta^{2}(1+\Lambda x) \,\bigr]
     - \alpha\ell_1 \bar R_2 - \beta\ell_2 \bar R_1 \,,
\end{equation}
a cubic. The radial part of \eqref{eq:dTsplit} is therefore a ratio of two cubics whose
only poles are the zeros of $\bar\Delta$, that is, by \eqref{eq:Deltabar}, the horizons.
The apparent singularity at $\Xir = 0$, present term by term in \eqref{eq:bar-tau}, is
absent from the sum.

The polar part is simpler. By \eqref{eq:linearinu}, $\Dth = \Xib + \Lambda\delta\,u$ is
linear in $u$ while $\Lambda\bar\chi = \Dth - 1$, so that
\begin{equation} \label{eq:polarsplit}
  \frac{\bar\chi}{\Dth} = \frac{1}{\Lambda}\left( 1 - \frac{1}{\Dth} \right) ,
  \qquad
  \Dth = \Lambda\delta\,\bigl( u - u_\Lambda \bigr) \,, \qquad
  u_\Lambda := -\,\frac{\Xib}{\Lambda\delta} \,,
\end{equation}
again a constant plus a single simple pole. That pole is never reached: on $u \in [0,1]$
the first of \eqref{eq:linearinu} exhibits $\Dth$ as a convex combination of $\Xia$ and
$\Xib$, so $\Dth > 0$ throughout the polar range whenever both are positive, which is the
condition for \eqref{eq:ds2} to be a regular metric in the first place.

Write $x_1, x_2, x_3$ for the zeros of the cubic $\bar\Delta$, so that
$\bar\Delta(x) = -\Lambda\,(x-x_1)(x-x_2)(x-x_3)$. These are the horizon radii in the
rescaled squared variable $x = \xi^{2}$, and any that are not real occur in conjugate
pairs. Since the leading coefficient of \eqref{eq:NTexplicit} is $\Xia\Xib\Ecal$, the
radial part of \eqref{eq:dTsplit} decomposes as
\begin{equation} \label{eq:pfT}
  \frac{\mathcal{N}_T(x)}{\bar\Delta(x)}
    = -\,\frac{\Xia\Xib\,\Ecal}{\Lambda}
      + \sum_{h=1}^{3} \frac{\mathcal{C}_h}{x - x_h} \,,
  \qquad
  \mathcal{C}_h := \frac{\mathcal{N}_T(x_h)}{\bar\Delta'(x_h)} \,,
\end{equation}
and the constant here is exactly minus the one delivered by \eqref{eq:polarsplit}. The two
cancel, leaving
\begin{equation} \label{eq:dTfinal}
  \frac{\dd T}{\dd\gamma}
    = \sum_{h=1}^{3} \frac{\mathcal{C}_h}{x(\gamma) - x_h}
      \;-\; \frac{c_\theta}{\Lambda}\;\frac{1}{u(\gamma) - u_\Lambda} \,,
  \qquad
  c_\theta := \frac{\Xia\Xib\,\Ecal}{\Lambda\delta} \,.
\end{equation}

This cancellation is worth a remark, because it decides which transcendents appear. Every
term of \eqref{eq:dTfinal} is a simple pole, so the whole of $T(\gamma)$ reduces to
integrals of the third kind. Had either constant survived, the quadrature would have
contained $\int x\,\dd\gamma$ instead, and which kind of transcendent that is depends on
the degree of $P$. By \eqref{eq:xsolturn} the poles of $x$ sit where $\wp$ attains
$B := P''(x_R)/6$. Substituting \eqref{eq:ginvP} then gives, for any zero $x_R$ of $P$, the
elementary identity
\begin{equation} \label{eq:branchvalue}
  4B^{3} - g_2^{(r)}B - g_3^{(r)} = 4\,p_4\,\bigl[\, P'(x_R) \,\bigr]^{2} \,.
\end{equation}
At a simple zero with $p_4 \neq 0$ the right-hand side is nonzero, so $B$ is not a branch
value of the lattice: $\wp - B$ has two simple zeros, $x$ has two simple poles, and
$\int x\,\dd\gamma$ would still be of the third kind. When the quartic degenerates to a
cubic the right-hand side vanishes, the two zeros merge into a double one, and
$\int x\,\dd\gamma$ becomes an elliptic integral of the second kind, contributing $\zeta$.
The same holds in the polar sector with $q_k$ in place of $p_k$. Since
$p_4 = -\Lambda\eps$ by \eqref{eq:p4}, a second-kind integral can arise at all only for
$\Lambda = 0$ or $\eps = 0$, and by the cancellation just made it in fact arises only in
the Myers--Perry limit, where $\bar\Delta$ drops to a quadratic and \eqref{eq:dTsplit}
becomes $\dd T/\dd\gamma = \Ecal\,\bar\rho^{2} + N/\bar\Delta$. That limit and the other
degenerations of Table \ref{tab:degrees} are taken up in Sec.~\ref{sec:degenerate}.

The residues of \eqref{eq:pfT} take a transparent form. At a zero of $\bar\Delta$ one has
$\Xir \bar R_1 \bar R_2 = x$, so that \eqref{eq:Ndef} and \eqref{eq:dTsplit} give
$\mathcal{N}_T(x_h) = N(x_h)/\Xir(x_h) = x_h \Psi(x_h)/\Xir(x_h)$. The identity
\eqref{eq:key} now collapses $\Psi$ exactly as it does in Appendix \ref{app:kg}: setting
\begin{equation} \label{eq:Omegadef}
  \Omega_i(x) := \frac{\alpha_i \Xii}{\bar R_i(x)} \,,
  \qquad\mbox{one has}\qquad
  \Xir(x)\,\Psi(x) = E - \sum_{i=1}^{2} \Omega_i(x)\,\ell_i
\end{equation}
identically in $x$, because $\Lambda + \Xir/\bar R_i = \Xii/\bar R_i$. Hence
\begin{equation} \label{eq:residue}
  \mathcal{C}_h
   = \frac{x_h}{\Xir(x_h)^{2}\,\bar\Delta'(x_h)}
     \left[\, E - \sum_{i=1}^{2} \Omega_i(x_h)\,\ell_i \,\right] .
\end{equation}
When $x_h$ is the event horizon, $\Omega_i(x_h)$ is its angular velocity in the
Boyer--Lindquist frame, the same quantity that appears in \eqref{eq:horexp}, and the
bracket is the superradiance combination $E - \Omega_i \ell_i$. The residue of the time
quadrature at a horizon is proportional to it, and vanishes identically at the superradiant
threshold.

What remains is to integrate a simple pole along the orbit, and here the two sectors pose
one and the same problem. Abbreviate
$\wp^{(r)}(z) := \wp\bigl(z;g_2^{(r)},g_3^{(r)}\bigr)$, and likewise $\wp^{(\theta)}$ for
the polar invariants. Take $x(\gamma)$ in the turning-point form \eqref{eq:xsolturn}, with
$x_R$ a zero of $P$ and $z = \gamma - \gamma_0$, and let $x_*$ be a constant that the orbit
does not reach. A one-line calculation moves the pole from $x$ to $\wp$,
\begin{equation} \label{eq:mobius}
  \frac{1}{x(\gamma) - x_*}
   = \frac{1}{x_R - x_*}
     - \frac{P'(x_R)}{(x_R-x_*)^{2}}\;\frac{1}{\wp^{(r)}(z) - \wp_*} \,,
  \qquad
  \wp_* := \frac{P''(x_R)}{6} - \frac{P'(x_R)}{x_R - x_*} \,,
\end{equation}
and $v_*$ denotes a point of the period parallelogram with $\wp^{(r)}(v_*) = \wp_*$.

The integral of $\bigl(\wp - \wp(v)\bigr)^{-1}$ is classical. With $\zeta$ and $\sigma$ the
Weierstrass functions of the same lattice, normalized by $\zeta' = -\wp$ and
$\sigma'/\sigma = \zeta$, put
\begin{equation} \label{eq:thirdkind}
  \Pi(z;v) := \frac{1}{\wp'(v)}
    \left[\; 2z\,\zeta(v) + \ln\frac{\sigma(v-z)}{\sigma(v+z)} \;\right] .
\end{equation}
Then $\Pi(0;v) = 0$ and
\begin{equation} \label{eq:thirdkindder}
  \frac{\partial \Pi}{\partial z}(z;v) = \frac{1}{\wp(z) - \wp(v)} \,,
\end{equation}
which follows from the addition theorem
$\zeta(z \pm v) = \zeta(z) \pm \zeta(v)
 + \half\bigl[\wp'(z) \mp \wp'(v)\bigr]/\bigl[\wp(z)-\wp(v)\bigr]$
on subtracting the two signs, since that gives
$\zeta(z-v) - \zeta(z+v) = -2\zeta(v) + \wp'(v)/\bigl[\wp(z)-\wp(v)\bigr]$. Expression
\eqref{eq:thirdkind} is the elliptic integral of the third kind in Weierstrass form, and it
is the only transcendent beyond $\wp$ itself that the problem requires.

Combining \eqref{eq:mobius} and \eqref{eq:thirdkindder} and doing the same in the polar
sector, where $u_R$ denotes a zero of $Q$, which by \eqref{eq:Qends} always exists, define
\begin{align}
  \mathcal{A}_r(x_*;z) &:= \frac{z}{x_R - x_*}
      - \frac{P'(x_R)}{(x_R-x_*)^{2}}\;\Pi^{(r)}(z;v_*) \,,
  \qquad
  \wp^{(r)}(v_*) = \frac{P''(x_R)}{6} - \frac{P'(x_R)}{x_R-x_*} \,,
  \label{eq:Arad}\\[4pt]
  \mathcal{A}_\theta(u_*;z) &:= \frac{z}{u_R - u_*}
      - \frac{Q'(u_R)}{(u_R-u_*)^{2}}\;\Pi^{(\theta)}(z;w_*) \,,
  \qquad
  \wp^{(\theta)}(w_*) = \frac{Q''(u_R)}{6} - \frac{Q'(u_R)}{u_R-u_*} \,,
  \label{eq:Apol}
\end{align}
where $\Pi^{(r)}$ and $\Pi^{(\theta)}$ are built from the invariants
$\bigl(g_2^{(r)},g_3^{(r)}\bigr)$ and $\bigl(g_2^{(\theta)},g_3^{(\theta)}\bigr)$ of
\eqref{eq:ginvP} respectively. By construction
$\dd\mathcal{A}_r/\dd\gamma = 1/(x-x_*)$ and
$\dd\mathcal{A}_\theta/\dd\gamma = 1/(u-u_*)$. The coefficient multiplying $\Pi$ has a
meaning of its own. The residue of $1/(x-x_*)$ at $z = v_*$ is
$-P'(x_R)/(x_R-x_*)^{2}$ divided by $\wp'(v_*)$. On the other hand
$\dd x/\dd\gamma = \pm 2\sqrt{P(x)}$, so that residue equals
$\pm 1/\bigl(2\sqrt{P(x_*)}\bigr)$. The sign records which of $\pm v_*$ has been taken.

The time coordinate is now explicit. Writing
$\bigl[\, g \,\bigr]_{\gamma_0}^{\gamma} := g(\gamma) - g(\gamma_0)$ and letting
$z_r = \gamma - \gamma_r$ and $z_\theta = \gamma - \gamma_\theta$ be measured from the Mino
times at which the two reference turning points $x_R$ and $u_R$ are attained, which need
not coincide, \eqref{eq:dTfinal} integrates to
\begin{equation} \label{eq:Tsol}
  T(\gamma) = T(\gamma_0)
    + \sum_{h=1}^{3} \mathcal{C}_h
      \Bigl[\, \mathcal{A}_r(x_h;z_r) \,\Bigr]_{\gamma_0}^{\gamma}
    - \frac{c_\theta}{\Lambda}
      \Bigl[\, \mathcal{A}_\theta(u_\Lambda;z_\theta) \,\Bigr]_{\gamma_0}^{\gamma} \,.
\end{equation}
Three comments are in order. First, \eqref{eq:Tsol} is real even though individual terms
need not be: zeros of $\bar\Delta$ that are not real occur in conjugate pairs, and so do
the associated $\mathcal{C}_h$ and $v_h$, so the two contributions add to twice a real part.
Second, the logarithm in \eqref{eq:thirdkind} must be followed continuously in $z$ rather
than evaluated on a principal branch. Crossing a branch cut shifts $\ln\sigma$ by $2\pi i$
and \eqref{eq:Tsol} by a spurious constant. In practice the argument is tracked along
the orbit, as in the implementations of \cite{Cieslik:2023qdc}. Third, $T$ is not periodic in
$\gamma$. Over one period of its lattice each $\mathcal{A}$ of
\eqref{eq:Arad}--\eqref{eq:Apol} increases by a fixed amount, from the term linear in $z$
and from the quasi-periodicity of $\sigma$. Hence $T$ advances by a fixed amount per
radial and per polar period while $x$ and $u$ merely oscillate. Coordinate time grows
secularly, as it must.

One case is left over. Equation \eqref{eq:mobius} uses the turning-point form
\eqref{eq:xsolturn}, hence a real zero of $P$, which by the discussion of
Sec.~\ref{sec:radialsol} may fail to exist when $\Lambda > 0$. Nothing essential changes.
For any solution of \eqref{eq:PQode}, $x(\gamma)$ is elliptic of order two, so
$1/(x-x_*)$ has exactly two simple poles in a period parallelogram, at the points
$z_*^{\pm}$ where $x = x_*$ is attained with $\dd x/\dd\gamma = \pm 2\sqrt{P(x_*)}$. Its
residues are $\pm 1/\bigl(2\sqrt{P(x_*)}\bigr)$ and therefore sum to zero, so that
\begin{equation} \label{eq:genericpole}
  \frac{1}{x(\gamma) - x_*}
   = c_* + \frac{\zeta\bigl(z - z_*^{+}\bigr) - \zeta\bigl(z - z_*^{-}\bigr)}
                {2\sqrt{P(x_*)}} \,,
\end{equation}
with $c_*$ fixed by evaluating at any convenient point. Integrating gives
\begin{equation} \label{eq:genericpoleint}
  \int \frac{\dd\gamma}{x(\gamma) - x_*}
   = c_*\,z + \frac{1}{2\sqrt{P(x_*)}}\,
     \ln\frac{\sigma\bigl(z - z_*^{+}\bigr)}{\sigma\bigl(z - z_*^{-}\bigr)} \,,
\end{equation}
of the same shape as \eqref{eq:Arad}. When a turning point is used,
$z_*^{\pm} = \pm v_*$ and the two expressions coincide.

%%--------------------------------------------------------------------- V.E --
\subsection{The azimuthal angles}\label{sec:azimuthalsol}

The two azimuthal equations are handled the same way, with one extra ingredient. Splitting
\eqref{eq:bar-phi} as before,
\begin{equation} \label{eq:dphisplit}
  \frac{\dd\ph_i}{\dd\gamma} = \frac{\mathcal{N}_i(x)}{\bar\Delta(x)}
     + \frac{\Xii \ell_i}{\mu_i^{2}}
     + \Lambda\alpha_i\,\Xia\Xib\,\Ecal\,\frac{\bar\chi(u)}{\Dth(u)} \,,
\end{equation}
with
\begin{equation} \label{eq:Nidef}
  \mathcal{N}_i := \frac{1}{\Xir \bar R_i}
    \Bigl[\, \Lambda\alpha_i \Xia\Xib\,\Ecal\,x\,\bar R_i\,\bar\Delta
           + \Xii\bigl(\alpha_j^{2}-\alpha_i^{2}\bigr)\ell_i\,\Xir\,\bar\Delta
           + \alpha_i \Xii\,N \,\Bigr] \,,
\end{equation}
where $j \neq i$ throughout. Now \emph{two} apparent poles have to disappear. That at
$\Xir = 0$ goes as in \eqref{eq:divXir}, the relevant identity being
$\bar R_j - \Lambda\Xij\,x^{2} = \Xir\bigl[\,x + \alpha_j^{2}(1+\Lambda x)\,\bigr]$. That at
$\bar R_i = 0$ is the cancellation already met below \eqref{eq:Qdef}: at
$x = -\alpha_i^{2}$ one has $\Xir = \Xii$, $\bar\Delta = \alpha_i^{2}$,
$\bar R_j = \alpha_j^{2}-\alpha_i^{2}$ and, from \eqref{eq:Ndef},
$N = -\Xii\alpha_i\ell_i(\alpha_j^{2}-\alpha_i^{2})$, so that the second and third terms of
\eqref{eq:Nidef} cancel while the first carries an explicit factor $\bar R_i$. Carrying out
both divisions,
\begin{equation} \label{eq:Niexplicit}
  \mathcal{N}_i = \Xii \Bigl\{\,
      \Ecal\,\alpha_i \bigl[\, \Lambda\Xij\,x\,\bar R_i \bar R_j
                             + x + \alpha_j^{2}(1+\Lambda x) \,\bigr]
    + \ell_i \bigl[\, (\alpha_j^{2}-\alpha_i^{2})\,\Xir \bar R_j
                    - \alpha_j^{2} \,\bigr]
    - \alpha_i\alpha_j \ell_j \,\Bigr\} \,,
\end{equation}
again a cubic, with leading coefficient $\Lambda\alpha_i\Xia\Xib\Ecal$. The poles of
\eqref{eq:dphisplit} in the radial variable are therefore once more the zeros of
$\bar\Delta$ and nothing else. This is the content of the observation of
\cite{Grunau:2017uwv} that the poles of their partial fractions are the zeros of
$\Delta_x$, and \eqref{eq:divXir} together with the evaluation at $x = -\alpha_i^{2}$ is
the reason for it.

The constant of the partial fraction is now $-\alpha_i\Xia\Xib\Ecal$, which is canceled by
the $\Lambda\alpha_i\Xia\Xib\Ecal/\Lambda$ coming from \eqref{eq:polarsplit}, exactly as in
the time equation. The residues follow from \eqref{eq:residue} without further work. At a
zero of $\bar\Delta$ the first two terms of \eqref{eq:Nidef} drop out, so that by
\eqref{eq:Omegadef}
\begin{equation} \label{eq:phires}
  \mathcal{N}_i(x_h) = \frac{\alpha_i \Xii\,N(x_h)}{\Xir(x_h)\,\bar R_i(x_h)}
     = \Omega_i(x_h)\,\mathcal{N}_T(x_h) \,,
  \qquad\mbox{whence}\qquad
  \frac{\mathcal{N}_i(x_h)}{\bar\Delta'(x_h)} = \Omega_i(x_h)\,\mathcal{C}_h \,.
\end{equation}
Finally $\mu_1^{2} = 1-u$ and $\mu_2^{2} = u$, so the centrifugal terms of
\eqref{eq:dphisplit} are simple poles at the two ends of the polar range. Collecting,
\begin{align}
  \frac{\dd\ph_1}{\dd\gamma}
    &= \sum_{h=1}^{3} \frac{\Omega_1(x_h)\,\mathcal{C}_h}{x - x_h}
     - \frac{\Xia \ell_1}{u-1} - \frac{\alpha\,c_\theta}{u - u_\Lambda} \,,
  \label{eq:dphi1final}\\[4pt]
  \frac{\dd\ph_2}{\dd\gamma}
    &= \sum_{h=1}^{3} \frac{\Omega_2(x_h)\,\mathcal{C}_h}{x - x_h}
     + \frac{\Xib \ell_2}{u} - \frac{\beta\,c_\theta}{u - u_\Lambda} \,,
  \label{eq:dphi2final}
\end{align}
and the quadratures are of the same three kinds as before. With the notation of
\eqref{eq:Arad}--\eqref{eq:Apol},
\begin{align}
  \ph_1(\gamma) &= \ph_1(\gamma_0)
    + \sum_{h=1}^{3} \Omega_1(x_h)\,\mathcal{C}_h
      \Bigl[\, \mathcal{A}_r(x_h;z_r) \,\Bigr]_{\gamma_0}^{\gamma}
    - \Xia \ell_1 \Bigl[\, \mathcal{A}_\theta(1;z_\theta) \,\Bigr]_{\gamma_0}^{\gamma}
    - \alpha\,c_\theta
      \Bigl[\, \mathcal{A}_\theta(u_\Lambda;z_\theta) \,\Bigr]_{\gamma_0}^{\gamma} \,,
  \label{eq:phi1sol}\\[4pt]
  \ph_2(\gamma) &= \ph_2(\gamma_0)
    + \sum_{h=1}^{3} \Omega_2(x_h)\,\mathcal{C}_h
      \Bigl[\, \mathcal{A}_r(x_h;z_r) \,\Bigr]_{\gamma_0}^{\gamma}
    + \Xib \ell_2 \Bigl[\, \mathcal{A}_\theta(0;z_\theta) \,\Bigr]_{\gamma_0}^{\gamma}
    - \beta\,c_\theta
      \Bigl[\, \mathcal{A}_\theta(u_\Lambda;z_\theta) \,\Bigr]_{\gamma_0}^{\gamma} \,,
  \label{eq:phi2sol}
\end{align}
which together with \eqref{eq:Tsol}, \eqref{eq:xsol} and \eqref{eq:usol} complete the
solution of the geodesic equations. Both $\mathcal{A}_\theta(0;\cdot)$ and
$\mathcal{A}_\theta(1;\cdot)$ are finite along the orbit, since by \eqref{eq:Qends} the
axes are never reached when the corresponding angular momentum is nonzero, and when it
vanishes the offending term disappears from \eqref{eq:phi1sol} or \eqref{eq:phi2sol}
altogether.

Three features of this pair deserve comment.

\begin{figure}[tbp]
  \centering
  \includegraphics[width=0.76\columnwidth]{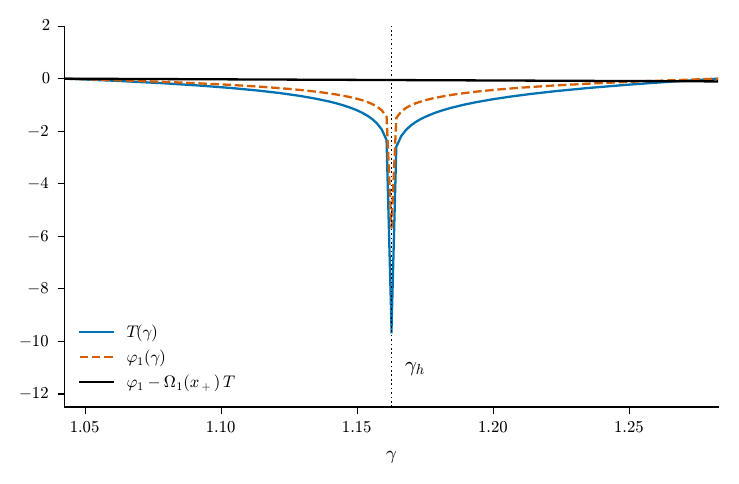}
  \caption{A horizon crossing, for $\alpha = 1/2$, $\beta = 3/10$, $\Lambda = -1/10$,
    $\eps = -1$, $E = 6/5$, $\ell_1 = 2/5$, $\ell_2 = 1/5$ and $\Kcal = 1/2$, computed
    from \eqref{eq:dTfinal} and \eqref{eq:dphi1final}. The orbit runs between
    $x = 0.0360$ and $x = 5.0380$ with radial period $2\omega_r = 4.7994$, and $\gamma$ is
    measured from the inner turning point, so that $x(0) = 0.0360$. It meets the event
    horizon $x_+ = 0.5667$ at the Mino time $\gamma_h = 1.1624$ marked by the dotted line,
    where $\Omega_1(x_+) = 0.5969$. Both $T$ and $\ph_1$ diverge
    logarithmically there and are defined only up to an additive constant on each side of
    the crossing, fixed here independently. The integrand of
    $\ph_1 - \Omega_1(x_+)\,T$ is regular at $x = x_+$ by \eqref{eq:phires}, and the
    combination passes through unchanged. Drawn with a Python script written with the assistance
    disclosed in the Acknowledgments; the values quoted were verified independently.}
  \label{fig:horizon}
\end{figure}

The first concerns the horizon. The radial logarithms enter $\ph_i$ and $T$ in the fixed
ratio $\Omega_i(x_h)$, so that in the combination $\ph_i - \Omega_i(x_h)\,T$ the entire
contribution of the zero $x_h$ cancels. For the event horizon this is the familiar
statement that a Boyer--Lindquist observer sees an infalling orbit wind about the hole
without limit while the co-rotating angle stays finite. Here it is visible at the level of
the closed-form solution rather than as an asymptotic estimate, and it holds separately for
each zero of $\bar\Delta$. Figure \ref{fig:horizon} shows the cancellation at an event
horizon.

The second concerns superradiance. By \eqref{eq:residue} the common factor
$\mathcal{C}_h$ carries the bracket \mbox{$E - \Omega_i(x_h)\ell_i$}. A geodesic whose
constants of motion sit at the superradiant threshold of a given horizon therefore receives
no contribution at all from that zero, in $T$ and in both angles simultaneously, and the
corresponding logarithm is absent from \eqref{eq:Tsol} and
\eqref{eq:phi1sol}--\eqref{eq:phi2sol}. This is the geodesic counterpart of the collapse of
the indicial exponent recorded in \eqref{eq:horexp}.

The third concerns infinity. The pole at $u = u_\Lambda$ enters $T$ with weight
$-c_\theta/\Lambda$ and $\ph_i$ with weight $-\alpha_i c_\theta$, hence in the ratio
$\Lambda\alpha_i$, so that it cancels from $\ph_i - \Lambda\alpha_i T$. That combination is
not accidental: $\dd\ph_i - \lambda a_i\,\dd\tau$ is precisely the one-form carried by the
last term of the metric \eqref{eq:ds2}, and at $M = 0$, where \eqref{eq:ds2} is
(anti-)de~Sitter space written in spheroidal coordinates, the substitution
$\hat\ph_i = \ph_i - \lambda a_i \tau$ is what diagonalizes it. Thus $\Lambda\alpha_i$ is
the angular velocity of the Boyer--Lindquist frame relative to one that does not rotate at
infinity, and the polar sector of the quadratures is governed by the asymptotic rotation in
the same way that the radial sector is governed by the rotation of the horizons.

It remains to set all of this beside \cite{Grunau:2017uwv}. Their Sec.~V\,C--E follows the
same route in a different order: the equations are rewritten using the radial and polar
equations, split into a radial and a polar integral, their Eq.~(46), each brought to
Weierstrass form by the substitutions of their Sec.~V\,A--B and then partial-fractioned,
their Eqs.~(47)--(49), (52) and (54). The resulting integrals are of the third kind, and
are solved in terms of $\wp$, $\sigma$ and $\zeta$ in their Eqs.~(50), (53) and (55). They record
that the poles correspond to the zeros of $\Delta_x$ and that the same poles occur in all
three equations. Our \eqref{eq:dTfinal} and
\eqref{eq:dphi1final}--\eqref{eq:dphi2final} have the same shape, and supply the reason the
poles are those of $\bar\Delta$ alone: the pole at $\Xir = 0$ is removed by
\eqref{eq:divXir} and the one at $\bar R_i = 0$ by the evaluation below \eqref{eq:Nidef}.

The comparison is nevertheless not term by term, and the two places where it is not are
instructive. The first is the azimuthal polar pole. Their Eqs.~(49) and (52) carry only the
poles at the two axes, $\nu = 1$ for $\phi$ and $\nu = 0$ for $\psi$, with no analogue of
our $u_\Lambda$, while their $t$ equation (54) does carry one. This is not a disagreement
but \eqref{eq:GNRangles}: their angles are ours less $\lambda a_i \tau$, which by the
remark above is exactly the combination from which the $\Dth$ pole is absent. Their polar
constants and poles are then the M\"obius reduction \eqref{eq:Apol} of $1/(u-1)$ and $1/u$,
with $q_\phi = Q''(u_R)/6 + Q'(u_R)/(1-u_R)$ and $q_\psi = Q''(u_R)/6 - Q'(u_R)/u_R$,
matching $\mathcal{A}_\theta(1;\cdot)$ and
$\mathcal{A}_\theta(0;\cdot)$. Correspondingly their radial residues are ours
with $\Omega_i(x_h)$ replaced by
$\Omega_i(x_h) - \Lambda\alpha_i = \alpha_i\Xir(x_h)/\bar R_i(x_h)$, the angular velocity of
the horizon measured in the frame that does not rotate at infinity. The second is where the
partial fraction is performed. Theirs is carried out after the passage to the Weierstrass
variable, so that their constants and residues refer to the chosen zeros $x_0$ and $\nu_0$.
Ours is carried out in $x$ and $u$, before that passage. The poles are then the $x_h$
themselves, and the residues \eqref{eq:residue} are independent of which turning point is
used and carry $\Omega_i$ and the superradiance bracket explicitly.

Three differences of substance remain. The residues are in closed form rather than left as
by-products of the decomposition, and with them the content of
Fig.~\ref{fig:horizon}: Sec.~VI of \cite{Grunau:2017uwv} observes that the orbits show
discontinuities at a horizon crossing, caused by divergences of the $t$, $\phi$ and $\psi$
equations there, and attributes them to the choice of coordinates. Equation
\eqref{eq:phires} identifies the combination in which they cancel, and \eqref{eq:residue}
shows that they are absent altogether at the superradiant threshold. The cancellation of the two constants in
\eqref{eq:dTfinal} shows that no elliptic integral of the second kind occurs when
$\Lambda \neq 0$ and identifies $\Lambda = 0$ as the one case in which it does. And
\eqref{eq:genericpole} covers the orbits without a radial turning point, for which the
substitution underlying \eqref{eq:mobius}, and with it the construction of
\cite{Grunau:2017uwv}, has no real starting point.

%%--------------------------------------------------------------------- V.F --
\subsection{The affine parameter}\label{sec:affine}

The Mino parameter is not an affine parameter, and for most purposes one wants the affine
one back. Undoing \eqref{eq:mino} in the rescaled variables of Sec.~\ref{sec:rescaled}
gives
\begin{equation} \label{eq:detadgamma}
  \frac{\dd\eta}{\dd\gamma} = \bar\rho^{2} = x + \bar\chi
    = x + \beta^{2} + \delta\,u \,,
\end{equation}
with $\eta$ proper time for $\eps = -1$ and an affine parameter along the ray for
$\eps = 0$. What this asks for are the quadratures of $x$ and of $u$ themselves, which are
exactly the two that cancelled out of $T$ and $\ph_i$ in Sec.~\ref{sec:timesol}. Here
nothing cancels them, so the integral $\int x\,\dd\gamma$ that was set aside there has to
be faced.

It is of the same kind, and the work is already done. Both \eqref{eq:xsolturn} and
\eqref{eq:usolturn} write their variable as a constant plus a multiple of
$1/(\wp - B)$, with
\begin{equation} \label{eq:Bdef}
  B_r := \frac{P''(x_R)}{6} \,, \qquad B_\theta := \frac{Q''(u_R)}{6} \,,
\end{equation}
so \eqref{eq:thirdkindder} integrates them at sight. Let $v_r$ and $w_r$ be points of the
respective period parallelograms with $\wp^{(r)}(v_r) = B_r$ and
$\wp^{(\theta)}(w_r) = B_\theta$. Then $\int x\,\dd\gamma = x_R z_r + P'(x_R)\,\Pi^{(r)}(z_r;v_r)$
and $\int u\,\dd\gamma = u_R z_\theta + Q'(u_R)\,\Pi^{(\theta)}(z_\theta;w_r)$, so that
\begin{equation} \label{eq:etasol}
  \eta(\gamma) = \eta(\gamma_0)
    + \bigl( x_R + \beta^{2} + \delta\,u_R \bigr)(\gamma-\gamma_0)
    + P'(x_R) \Bigl[\, \Pi^{(r)}(z_r;v_r) \,\Bigr]_{\gamma_0}^{\gamma}
    + \delta\,Q'(u_R) \Bigl[\, \Pi^{(\theta)}(z_\theta;w_r) \,\Bigr]_{\gamma_0}^{\gamma} \,,
\end{equation}
in the notation of \eqref{eq:Tsol}. The affine parameter is therefore of the same third
kind as the time coordinate, and like it advances secularly while $x$ and $u$ merely
oscillate.

The coefficients standing in front of the logarithms are not new constants. Evaluating the
branch-value identity \eqref{eq:branchvalue} at $B_r$ gives
$\bigl[\wp^{(r)}{}'(v_r)\bigr]^{2} = 4B_r^{3} - g_2^{(r)}B_r - g_3^{(r)}
 = 4p_4\bigl[P'(x_R)\bigr]^{2}$, and $q_4 = \delta^{2}p_4$ by \eqref{eq:p4} and
\eqref{eq:q4}, so that
\begin{equation} \label{eq:etacoeff}
  \frac{P'(x_R)}{\wp^{(r)}{}'(v_r)} = \pm\,\frac{1}{2\sqrt{p_4}} \,, \qquad
  \frac{Q'(u_R)}{\wp^{(\theta)}{}'(w_r)}
     = \pm\,\frac{1}{2\sqrt{q_4}} = \pm\,\frac{1}{2\,|\delta|\sqrt{p_4}} \,,
\end{equation}
the sign recording which of $\pm v_r$ has been taken, as below \eqref{eq:Apol}. These
numbers have a direct reading. Near a pole of $x$ the equation
$(\dd x/\dd\gamma)^{2} = 4P(x)$ is dominated by its leading term, $4p_4x^{4}$, so that
$x \sim \mp 1/\bigl[2\sqrt{p_4}\,(\gamma-\gamma_\infty)\bigr]$: the constants
\eqref{eq:etacoeff} are the residues of $x$ and of $u$ at their own poles. Since those
residues sum to zero, as they must for an elliptic function, \eqref{eq:etasol} can be
written without reference to a turning point at all,
\begin{equation} \label{eq:etageneric}
  \int x(\gamma)\,\dd\gamma
   = c_\infty\,z \;\pm\; \frac{1}{2\sqrt{p_4}}\,
     \ln\frac{\sigma\bigl(z - z_\infty^{+}\bigr)}{\sigma\bigl(z - z_\infty^{-}\bigr)} \,,
\end{equation}
where $z_\infty^{\pm}$ are the two points of the period parallelogram at which $x$
diverges, the sign is fixed by which of them is labelled $z_\infty^{+}$, and $c_\infty$ is
fixed by evaluating at any convenient point. This is the
counterpart of \eqref{eq:genericpoleint} and covers, in the same way, the orbits of
Sec.~\ref{sec:radialsol} that have no radial turning point.

One case is genuinely different. By \eqref{eq:branchvalue} the right-hand side vanishes
when $p_4 = 0$, so $B_r$ is then a branch value, the two poles of $x$ merge into one double
pole, and $\int x\,\dd\gamma$ drops from the third kind to the second. By \eqref{eq:p4}
this is the case for null geodesics and in the Myers--Perry limit, where $P$ is a cubic and
the affine reduction \eqref{eq:cubicaffine} replaces \eqref{eq:xsolturn}. There
\eqref{eq:etasol} is superseded by
\begin{equation} \label{eq:etaMP}
  \eta(\gamma) = \eta(\gamma_0) + \beta^{2}(\gamma-\gamma_0)
    + \bigl[\, \mathcal{Z}_r(z_r + \omega_r) \,\bigr]_{\gamma_0}^{\gamma}
    + \delta \bigl[\, \mathcal{Z}_\theta(z_\theta + \omega_\theta)
               \,\bigr]_{\gamma_0}^{\gamma} \,,
\end{equation}
with $\mathcal{Z}_r$ and $\mathcal{Z}_\theta$ the second-kind functions \eqref{eq:Zdef} of
Sec.~\ref{sec:degenerate} and $\omega_r$, $\omega_\theta$ the half-periods defined there,
at which $\wp^{(r)}$ and $\wp^{(\theta)}$ attain $B_r$ and $B_\theta$. The elliptic integral of the second kind that the cancellation
of Sec.~\ref{sec:timesol} kept out of $T$ and $\ph_i$ thus reappears here, and the affine
parameter is where it belongs.

%%--------------------------------------------------------------------- V.G --
\subsection{The degenerate cases}\label{sec:degenerate}

Sections \ref{sec:radialsol} and \ref{sec:polarsol} already cover every row of
Table \ref{tab:degrees}, since neither \eqref{eq:xsol} nor \eqref{eq:usol} degenerates when
the leading coefficient of the quartic vanishes. The quadratures of
Secs.~\ref{sec:timesol} and \ref{sec:azimuthalsol} are another matter: they were written
for $\Lambda \neq 0$ and $\alpha \neq \beta$, and the two coefficients $c_\theta$ and
$u_\Lambda$ of \eqref{eq:polarsplit} and \eqref{eq:dTfinal} are singular in either limit.
We take the remaining rows in turn.

\paragraph*{Null geodesics, $\eps = 0$.}
Nothing changes. None of $\bar\Delta$, $N$, $\mathcal{N}_T$, $\mathcal{N}_i$, $\Omega_i$,
$\mathcal{C}_h$, $u_\Lambda$ and $c_\theta$ contains $\eps$: the causal character enters the
problem only through $P$ and $Q$, hence only through the invariants \eqref{eq:ginvP} and the
turning points. Equations \eqref{eq:Tsol} and \eqref{eq:phi1sol}--\eqref{eq:phi2sol}
therefore hold verbatim, with $P$ and $Q$ cubics. One point deserves notice. By
\eqref{eq:branchvalue} the vanishing of $p_4$ makes $B$ a branch value, so that
$x(\gamma)$ acquires a double pole. But what is integrated is $1/(x-x_*)$, whose $\wp_*$ in
\eqref{eq:mobius} is displaced from $B$ by $P'(x_R)/(x_R-x_*)$ and is therefore not a
branch value. Neither $\mathcal{A}_r$ nor $\mathcal{A}_\theta$ is affected. The affine
parameter is the one quantity that is: its integrand is $x$ itself, whose $\wp_*$ is $B$
and nothing else, so \eqref{eq:etasol} gives way to the second-kind form \eqref{eq:etaMP},
with $P$ and $Q$ the cubics of this limit.

\paragraph*{Myers--Perry, $\Lambda = 0$.}
Three things change. First, $\Xir = \Xia = \Xib = 1$ and $\Dth = 1$, so the polar simple
pole disappears altogether: $c_\theta$ and $u_\Lambda$ both diverge, but
$\bar\chi = \beta^{2} + \delta u$ now enters \eqref{eq:dTsplit} directly and
\eqref{eq:dphisplit} not at all. Second, $\bar\Delta = \bar R_1 \bar R_2 - x$ is a
\emph{quadratic}, so the sums over $h$ run over its two zeros, the Myers--Perry horizons.
The residue formula \eqref{eq:residue} survives unchanged, with $\Xir = 1$ and
$\Omega_i(x) = \alpha_i/\bar R_i$. Third, $\mathcal{N}_T$ remains a cubic while
$\mathcal{N}_i$ drops to degree one, so that the time equation acquires a polynomial part
and the azimuthal equations do not,
\begin{align}
  \frac{\dd T}{\dd\gamma} &= E\,\bigl( \bar\rho^{2} + 1 \bigr)
     + \sum_{h=1}^{2}\frac{\mathcal{C}_h}{x - x_h} \,,
  \label{eq:dTMP}\\[4pt]
  \frac{\dd\ph_i}{\dd\gamma} &= \sum_{h=1}^{2}
      \frac{\Omega_i(x_h)\,\mathcal{C}_h}{x - x_h} + \frac{\ell_i}{\mu_i^{2}} \,.
  \label{eq:dphiMP}
\end{align}

The new ingredient is $\int \bar\rho^{2}\,\dd\gamma$. Both $P$ and $Q$ are cubics here, and
for a cubic the classical reduction is affine rather than M\"obius: with the invariants
\eqref{eq:ginvP} unchanged,
\begin{equation} \label{eq:cubicaffine}
  x(\gamma) = \frac{\wp^{(r)}(z_r + \omega_r) - p_2/3}{p_3} \,, \qquad
  u(\gamma) = \frac{\wp^{(\theta)}(z_\theta + \omega_\theta) - q_2/3}{q_3} \,,
\end{equation}
as one verifies directly from \eqref{eq:PQode}. The shifts are not optional, and
\eqref{eq:branchvalue} is what supplies them. Setting $x = x_R$ in \eqref{eq:cubicaffine}
asks for $\wp^{(r)} = p_3x_R + p_2/3$, which for a cubic is $P''(x_R)/6 = B_r$, the very
quantity of \eqref{eq:Bdef}. At $p_4 = 0$ the right-hand side of \eqref{eq:branchvalue}
vanishes, so $B_r$ is a branch value and the point at which $\wp^{(r)}$ attains it is a
half-period: that half-period is $\omega_r$, and $\omega_\theta$ is its polar counterpart
with $\wp^{(\theta)}(\omega_\theta) = B_\theta$. The shift moves the pole of $\wp^{(r)}$, at
which \eqref{eq:cubicaffine} sends $x$ to infinity, from $z_r = 0$ to $z_r = -\omega_r$. The
origin $z_r = 0$ then lands on $x_R$, because $\wp^{(r)}(\omega_r) = B_r$. The sign of
$\omega_r$ is immaterial, since $\wp$ has period $2\omega_r$, so that $z_r = +\omega_r$ is
the same pole. Without
the shift \eqref{eq:cubicaffine} describes a different branch of the same curve, one that
does not pass through $x_R$ at all. Since $\zeta' = -\wp$, the two functions
\begin{equation} \label{eq:Zdef}
  \mathcal{Z}_r(z) := -\,\frac{\zeta^{(r)}(z) + p_2 z/3}{p_3} \,, \qquad
  \mathcal{Z}_\theta(z) := -\,\frac{\zeta^{(\theta)}(z) + q_2 z/3}{q_3} \,,
\end{equation}
satisfy $\dd\mathcal{Z}_r(z_r+\omega_r)/\dd\gamma = x$ and
$\dd\mathcal{Z}_\theta(z_\theta+\omega_\theta)/\dd\gamma = u$, carrying the same shifts:
these are the elliptic integrals of the second kind anticipated in
Sec.~\ref{sec:timesol}. With $\bar\rho^{2} = x + \beta^{2} + \delta u$,
\begin{equation} \label{eq:TMP}
  T(\gamma) = T(\gamma_0)
    + E \Bigl\{ \bigl( 1+\beta^{2} \bigr)(\gamma - \gamma_0)
       + \bigl[\, \mathcal{Z}_r(z_r + \omega_r) \,\bigr]_{\gamma_0}^{\gamma}
       + \delta \bigl[\, \mathcal{Z}_\theta(z_\theta + \omega_\theta)
                  \,\bigr]_{\gamma_0}^{\gamma} \Bigr\}
    + \sum_{h=1}^{2}\mathcal{C}_h
       \bigl[\, \mathcal{A}_r(x_h;z_r) \,\bigr]_{\gamma_0}^{\gamma} \,,
\end{equation}
while \eqref{eq:phi1sol} and \eqref{eq:phi2sol} survive with $\Xia = \Xib = 1$, with the
sums truncated to $h = 1,2$ and with the $u_\Lambda$ terms deleted. The affine parameter of
Sec.~\ref{sec:affine} uses the same two functions, as \eqref{eq:etaMP} records, which is
only to say that $\int\bar\rho^{2}\,\dd\gamma$ is what both of them need.

\paragraph*{Equal rotations, $\alpha = \beta$.}
Here $\delta = 0$, and $c_\theta$ and $u_\Lambda$ diverge together while their product stays
finite. What happens is that $\Dth = \Xi$ and $\bar\chi = \alpha^{2}$ become constants,
where $\Xi := \Xia = \Xib$, so that the polar pole of \eqref{eq:dTfinal} collapses to the
constant $-\Xi\Ecal/\Lambda$ and that of
\eqref{eq:dphi1final}--\eqref{eq:dphi2final} to $-\alpha\Xi\Ecal$. The radial sector is
untouched, $P$ remaining a quartic and $\bar\Delta$ a cubic. Writing
$\bar R := x+\alpha^{2}$ and $\Omega(x) := \alpha\Xi/\bar R$, which is now common to the two
planes,
\begin{align}
  \frac{\dd T}{\dd\gamma} &= -\,\frac{\Xi\,\Ecal}{\Lambda}
     + \sum_{h=1}^{3}\frac{\mathcal{C}_h}{x-x_h} \,,
  \label{eq:dTeq}\\[4pt]
  \frac{\dd\ph_i}{\dd\gamma} &= -\,\alpha\,\Xi\,\Ecal
     + \sum_{h=1}^{3}\frac{\Omega(x_h)\,\mathcal{C}_h}{x-x_h}
     + \frac{\Xi \ell_i}{\mu_i^{2}} \,.
  \label{eq:dphieq}
\end{align}
The time coordinate thus loses its polar quadrature entirely, and both $T$ and $\ph_i$
acquire a term linear in the Mino parameter. Since the radial contributions to $\ph_1$ and
$\ph_2$ have become identical, the two angles differ only through their centrifugal terms.
The affine parameter loses its polar part along with them: $\delta = 0$ makes
$\bar\rho^{2} = x + \alpha^{2}$, so that only the radial term of \eqref{eq:etasol}
survives.

The polar sector itself turns elementary, which is what the last two rows of
Table \ref{tab:degrees} record. With $q_4 = q_3 = 0$ the second of \eqref{eq:PQode} is
solved by a sinusoid, the origin of $z_\theta$ now sitting at the upper of the two polar
turning points, $u_R = \bar u + A$,
\begin{equation} \label{eq:usin}
  u(\gamma) = \bar u + A\cos\bigl( \omega z_\theta \bigr) \,, \qquad
  \bar u = -\frac{q_1}{2q_2} \,, \qquad
  A = \frac{\sqrt{q_1^{2}-4q_0q_2}}{2\,|q_2|} \,, \qquad
  \omega = 2\sqrt{-q_2} \,,
\end{equation}
and $\mathcal{A}_\theta$ of \eqref{eq:Apol} is replaced by
\begin{equation} \label{eq:Aelem}
  \mathcal{A}_\theta(u_*;z) = \frac{2}{\omega\,(D-A)\,k}\,
     \arctan\!\left[ \frac{1}{k}\,\tan\frac{\omega z}{2} \right] ,
  \qquad D := \bar u - u_* \,, \qquad
  k := \sqrt{\frac{D+A}{D-A}} \,,
\end{equation}
which is real at $u_* = 0$ and $u_* = 1$ because $Q(0)$ and $Q(1)$ are negative by
\eqref{eq:Qends}, so that $|D| > A$ and $D \pm A$ carry the same sign. As with the
logarithm in \eqref{eq:thirdkind}, the arctangent is to be followed continuously rather
than taken on its principal branch.

\paragraph*{$\Lambda = 0$ and $\alpha = \beta$.}
The two sets of changes superpose. Now $\bar\chi = \alpha^{2}$ is constant and
$\bar\Delta = \bar R^{2} - x$ is a quadratic, so that
\begin{equation} \label{eq:MPeq}
  \frac{\dd T}{\dd\gamma} = E\bigl( x + \alpha^{2} + 1 \bigr)
     + \sum_{h=1}^{2}\frac{\mathcal{C}_h}{x-x_h} \,,
  \qquad
  \frac{\dd\ph_i}{\dd\gamma} = \sum_{h=1}^{2}
     \frac{\Omega(x_h)\,\mathcal{C}_h}{x-x_h} + \frac{\ell_i}{\mu_i^{2}} \,,
\end{equation}
with $\Omega(x) = \alpha/\bar R$. The time coordinate now needs $\mathcal{Z}_r$ of
\eqref{eq:Zdef} but no $\mathcal{Z}_\theta$, the polar motion having dropped out of
$\bar\rho^{2}$, and the angles need only the two radial integrals of the third kind
together with the elementary \eqref{eq:Aelem}. The affine parameter reduces with them, to
$\eta = \eta(\gamma_0) + \alpha^{2}(\gamma-\gamma_0)
 + [\,\mathcal{Z}_r(z_r+\omega_r)\,]_{\gamma_0}^{\gamma}$. Every row of Table \ref{tab:degrees} is
thus accounted for in closed form.

%%%%%%%%%%%%%%%%%%%%%%%%%%%%%%%%%%%%%%%%%%%%%%%%%%%%%%%%%%%%%%%%%%%%%%%%%%%%%%
%%  VI.  CONCLUSIONS AND OUTLOOK
%%%%%%%%%%%%%%%%%%%%%%%%%%%%%%%%%%%%%%%%%%%%%%%%%%%%%%%%%%%%%%%%%%%%%%%%%%%%%%
\section{Conclusions and outlook}\label{sec:conclusions}

We have gone back over the geodesic problem of the five-dimensional Kerr--(A)dS black hole
in Boyer--Lindquist coordinates and carried it through to closed-form solutions. Four things
came out of it.

The first is a correction. One block of the inverse metric of Sec.~\ref{sec:inverse}
disagrees with the expression quoted in \cite{Kunduri:2005fq}, and the radial potential that the
Hamilton--Jacobi separation produces is not the $\widetilde{V}$ of Eq.~(16) of that
reference but \eqref{eq:Vtilde}. The discrepancy \eqref{eq:Vexcess} is a genuine function of
$r$ and $\theta$ and not a matter of normalization. It matters. As shown in
Appendix \ref{app:kg}, the excess has a simple pole at the horizon. The indicial exponents
of the radial Klein--Gordon equation are therefore shifted by an $O(1)$ amount, and with
them the ingoing boundary condition and the superradiance criterion. The corrected
potential returns the combination $\omega - \Omega_i m_i$ automatically, as it must, while
that of Eq.~(16) of \cite{Kunduri:2005fq} does not. Any computation of quasinormal modes or of superradiant
amplification in this background that started from that equation should be redone.

The second is a matter of bookkeeping that turns out to carry weight. The separation constant
is fixed at the step where it first appears, through the pole-free splitting of
Sec.~\ref{sec:ids}, rather than afterwards. Of the three normalizations now in circulation
only $\Kkl$ fails to have a limit as $\lambda \to 0$. Ours and that of
\cite{Grunau:2017uwv} differ from each other at $O(\lambda)$, and both reduce to the
constant of \cite{Frolov:2003en}.

The third is the degree statement of Sec.~\ref{sec:polynomials}. In the dimensionless
variables of Sec.~\ref{sec:rescaled} both orbital polynomials are quartic, with leading
coefficients proportional to $\Lambda\eps$, and Table \ref{tab:degrees} collects the
degenerations. Set beside Table \ref{tab:degrees4d} this is the substantive difference
between four and five dimensions: because the natural radial variable is $x = r^{2}$, the
cosmological constant raises the radial degree only from three to four, whereas in
Kerr--(A)dS it raises it from four to six and forces the Kleinian $\sigma$ functions of
\cite{Hackmann:2010zz}. Both sectors stay elliptic in five dimensions, for timelike
geodesics as well as null.

The fourth is the solution itself. Section \ref{sec:bw} carries the
Biermann--Weierstrass construction of
\cite{Cieslik:2022uki,Cieslik:2023qdc} over to both sectors at once. Equations
\eqref{eq:xsol} and \eqref{eq:usol} are a single pair of formulas, manifestly real,
parametrized by the constants of motion and by an arbitrary point of the orbit rather than
by a turning point, and valid without case splitting for every row of
Table \ref{tab:degrees}. The freedom in the starting point is not a convenience: for
$\Lambda > 0$ the radial quartic can be positive for every real $x$, as in
Fig.~\ref{fig:nots}, and the orbit then has no turning point at which the classical
substitution could be started at all. The quadratures for $T$ and $\ph_i$ reduce to elliptic
integrals of the third kind alone, Eqs.~\eqref{eq:Tsol} and
\eqref{eq:phi1sol}--\eqref{eq:phi2sol}, and the affine parameter follows from
\eqref{eq:etasol} by one more of the same kind. Two structural facts make that work. The apparent
poles at $\Xir = 0$ and at $\bar R_i = 0$ cancel, leaving the zeros of $\bar\Delta$, i.e. the
horizons, as the only ones. And the constants of the radial and the polar partial fraction
cancel against each other, so that no integral of the second kind occurs unless
$\Lambda = 0$. The residues are explicit. By \eqref{eq:residue} and \eqref{eq:phires} they
are proportional to the superradiance bracket $E - \Omega_i(x_h)\ell_i$, and they stand in
the fixed ratio $\Omega_i(x_h)$ between the azimuthal and the temporal equation. This
accounts in closed form for the coordinate divergences at a horizon crossing noted in
Sec.~VI of \cite{Grunau:2017uwv}. They cancel in $\ph_i - \Omega_i(x_h)\,T$, as in
Fig.~\ref{fig:horizon}, and are absent altogether at the superradiant threshold.

Several directions suggest themselves.

The de~Sitter orbits are open. Reference \cite{Grunau:2017uwv} classifies the
anti-de~Sitter case in detail. For $\Lambda > 0$ there is a cosmological horizon, a third
positive zero of $\bar\Delta$, and the turning-point-free orbits of Fig.~\ref{fig:nots}.
Section \ref{sec:bw} supplies the solutions a corresponding classification would be built
on.

Observables follow from the periods. The radial and polar motions are elliptic with periods
$2\omega_r$ and $2\omega_\theta$, whose ratio is in general irrational, and $T$ and
$\ph_i$ advance by fixed amounts over each. The periastron shift, the Lense--Thirring precession of the two
rotation planes and the deflection of light are therefore read off from the terms linear in
$z$ in \eqref{eq:Arad}--\eqref{eq:Apol} together with the quasi-periods of $\zeta$ and
$\sigma$, and without the case splitting that usually accompanies them.

Higher dimensions pose the obvious structural question, and the answer is sharper than a
count of degrees. Integrability is not the issue. The general Kerr--NUT--(A)dS spacetime
carries a principal tensor, that is a non-degenerate closed conformal Killing--Yano
two-form, in every dimension, and the Killing tower it generates makes the geodesic motion
completely integrable and the Hamilton--Jacobi equation separable in all $D$
\cite{Page:2006ka,Frolov:2006pe,Frolov:2017kze}. Nor could the temporal and the azimuthal
sector be where anything broke down, since those coordinates are cyclic and enter the
separated action linearly, as in \eqref{eq:Ssol}, whatever $D$ may be.

What is particular to $D \leq 5$ is the rest of the route taken here. Separation in general
$D$ takes place in the canonical coordinates built from the eigenvalues of the principal
tensor, and not in Boyer--Lindquist-type ones: the Myers--Perry direction cosines carry the
constraint $\sum_i \mu_i^{2} = 1$, which makes them inconvenient for separation in more
than five dimensions \cite{Frolov:2017kze}. In five dimensions that constraint is solved
outright by the single angle $\theta$ of Sec.~\ref{sec:metricfns}, which is why
\eqref{eq:ds2} gives no trouble. More decisive is the Mino parameter. Written in the
canonical coordinates, the velocities of the $n$ non-cyclic variables $x_\mu$ carry factors
$U_\mu = \prod_{\nu \neq \mu}\bigl( x_\nu^{2} - x_\mu^{2} \bigr)$ that couple them exactly
as $\rhs$ couples the two equations \eqref{eq:drdtheta}, and these agree up to a sign only
for $n = 2$, which is to say only in $D = 4$ and $D = 5$. A single reparametrization
therefore decouples the sectors in those two cases and in no other \cite{Frolov:2017kze}.
For $n \geq 3$ what takes the place of \eqref{eq:PQode} is a Jacobi inversion problem in
$n$ variables, solved by Abelian functions of $n$ arguments rather than by elliptic
functions of one, and the polynomials under the square roots grow in degree with $D$
besides. Five dimensions is thus not merely the last case in which the orbital polynomials
stay quartic. It is the last case in which the four-dimensional strategy of
\cite{Cieslik:2023qdc} applies at all.

Finally there is the charged case, where part of the answer is already in. A charged
generalization of the present spacetime is known in minimal five-dimensional gauged
supergravity \cite{Chong:2005hr}, and the question raised here is whether
\eqref{eq:PQode} survives the addition of charge with both polynomials still of degree at
most four. For the supersymmetric Gutowski--Reall member, which carries two equal rotation
parameters, Drawer and Grunau \cite{Drawer:2020mpw} show that it does: their radial
polynomial is quartic with leading coefficient proportional to $\Lambda\eps$, as in
\eqref{eq:p4}, and their polar polynomial is quadratic, which is the $\alpha = \beta$ row
of Table \ref{tab:degrees}. Two differences from the treatment given here matter. Their
radial solution is the classical substitution about a zero of the quartic, their Eq.~(42),
and so inherits the restriction of \eqref{eq:gnrsol}, where \eqref{eq:xsol} needs no such
zero. And their horizon is degenerate, so their time quadrature carries a second-order
pole, their Eqs.~(52)--(53), where \eqref{eq:pfT} is written for simple zeros. Three things
remain to be done: the extremal limit of \eqref{eq:Tsol}, which needs that same double
pole; the rest of the charged family, meaning unequal rotation parameters and the general
non-supersymmetric solution of \cite{Chong:2005hr}; and charged test particles, which
\cite{Drawer:2020mpw} does not treat and \cite{Grunau:2017uwv} proposes.

%%%%%%%%%%%%%%%%%%%%%%%%%%%%%%%%%%%%%%%%%%%%%%%%%%%%%%%%%%%%%%%%%%%%%%%%%%%%%%
%%  ACKNOWLEDGMENTS
%%%%%%%%%%%%%%%%%%%%%%%%%%%%%%%%%%%%%%%%%%%%%%%%%%%%%%%%%%%%%%%%%%%%%%%%%%%%%%
%%  APS asks for the disclosure of substantive AI use in the Acknowledgments,
%%  and for any AI use in figure generation to be stated in the figure captions
%%  as well; see "Appropriate Use of AI Tools".
\begin{acknowledgments}
{\"O}.~S.~is grateful to Ozan Tu\u{g}luk for his intangible support.

The preparation of this work was assisted by the artificial-intelligence tool Claude
(Anthropic), version Claude Opus 5. It was used for symbolic and numerical algebra, for the
independent re-derivation and cross-checking of the identities, closed forms and polynomial
coefficients reported here, for the Python scripts that produce
Figs.~\ref{fig:nots}--\ref{fig:horizon}, and for drafting and revising the text. The
authors formulated the problem, directed the work throughout, and checked every statement,
every equation and every numerical value quoted against their own derivations and against
independent computation. The tool is not an author, and the authors take full
responsibility for the content of this paper.
\end{acknowledgments}

%%%%%%%%%%%%%%%%%%%%%%%%%%%%%%%%%%%%%%%%%%%%%%%%%%%%%%%%%%%%%%%%%%%%%%%%%%%%%%
%%  DATA AVAILABILITY STATEMENT, required by the Physical Review journals
%%%%%%%%%%%%%%%%%%%%%%%%%%%%%%%%%%%%%%%%%%%%%%%%%%%%%%%%%%%%%%%%%%%%%%%%%%%%%%
\section*{Data availability}

No data were created or analyzed in this study. Every result reported here is analytic and
is given in full in the text. The symbolic and numerical scripts used to verify the
identities and closed forms quoted above, and to produce
Figs.~\ref{fig:nots}--\ref{fig:horizon}, are available from the authors upon reasonable
request.

%%%%%%%%%%%%%%%%%%%%%%%%%%%%%%%%%%%%%%%%%%%%%%%%%%%%%%%%%%%%%%%%%%%%%%%%%%%%%%
%%  APPENDIX A.  THE KLEIN-GORDON EQUATION
%%%%%%%%%%%%%%%%%%%%%%%%%%%%%%%%%%%%%%%%%%%%%%%%%%%%%%%%%%%%%%%%%%%%%%%%%%%%%%
\appendix
\section{Consequences for the Klein--Gordon equation}\label{app:kg}

For completeness we record how the Klein--Gordon analysis in the later part of
\cite{Kunduri:2005fq} is affected. Almost all of it stands. Its reduction of the wave
operator, its angular equation, its Fuchsian form in $z = \sin^{2}\theta$ and its
hypergeometric and Jacobi-polynomial solutions in the degenerate case $a=b$, that is
Eqs.~(21)--(26) of that reference, are all unaffected. The reason is that the
only part of $g^{\ph_i\ph_j}$ entering the angular sector is the $\delta^{ij}$ term, through
the identity \eqref{eq:id13}, exactly as for the geodesic equation \eqref{eq:theta-eq}.
They are written with the separation constant of \cite{Kunduri:2005fq}, however, and we
restore ours below.

The radial equation, Eq.~(27) of \cite{Kunduri:2005fq}, is unaltered in form,
\begin{equation} \label{eq:KGrad}
  -\frac{1}{r}\frac{\dd}{\dd r}\!\left( r(V-2M)\frac{\dd R}{\dd r} \right)
  + \widetilde{V}(r;\omega,m_i,-\mu^{2})\,R = k R \,,
\end{equation}
with $E \to \omega$, $L_i \to m_i$, $\eps \to -\mu^{2}$ and $k = -K$, but $\widetilde{V}$
must now be the corrected function \eqref{eq:Vtilde}. Here $\mu$ is the mass of the scalar
field and $m_i \in \mathbb{Z}$ are the azimuthal quantum numbers, following the notation of
\cite{Kunduri:2005fq}. The choice $k = -K$ gives \eqref{eq:KGrad} and \eqref{eq:r-eq} the
same right-hand side, and the two then agree term by term. The only replacement is
\begin{equation} \label{eq:quantization}
  (V-2M)\left(\frac{\dd S_r}{\dd r}\right)^{\!2}
  \;\longrightarrow\;
  -\frac{1}{r}\frac{\dd}{\dd r}\!\left( r(V-2M)\frac{\dd}{\dd r} \right) ,
\end{equation}
which is $p_r \to -i\,\dd/\dd r$ carried out with the invariant measure
$\sqrt{-g}\,g^{rr}$. The minus sign it brings comes from the factor $i$, so
$\widetilde{V}$ keeps the sign it has in \eqref{eq:r-eq}. The eigenvalue and the potential
are shifted by one and the same
constant in passing from $\Kkl$ to $K$, so \eqref{eq:KGrad} holds in either normalization,
and the local analysis that follows (which depends on $\widetilde{V}$ only through its
residue at $r_+$) is insensitive to the choice.

The same replacement in the polar sector, with $\sqrt{-g}\,g^{\theta\theta}$ in place of
$\sqrt{-g}\,g^{rr}$, turns \eqref{eq:theta-eq} into the angular equation,
\begin{equation} \label{eq:KGang}
  -\frac{1}{\sin\theta\cos\theta}\frac{\dd}{\dd\theta}\!
    \left( \Dth \sin\theta\cos\theta\,\frac{\dd\Theta}{\dd\theta} \right)
  - \left( \eps + \frac{\Xia\Xib\,\Ecal^{2}}{\Dth} \right) \chit\,\Theta
  + \left( \frac{\Xia m_1^{2}}{\sin^{2}\theta}
         + \frac{\Xib m_2^{2}}{\cos^{2}\theta} \right) \Theta
  = K\,\Theta \,,
\end{equation}
which is Eq.~(22) of \cite{Kunduri:2005fq} with the constant of \eqref{eq:KvsKKL} moved,
exactly as for the geodesic pair. Neither \eqref{eq:KGrad} nor \eqref{eq:KGang} has a pole
at $\lambda = 0$, while Eq.~(22) of \cite{Kunduri:2005fq} and the potential of its Eq.~(16)
both do. The Myers--Perry limit of the Klein--Gordon equations is therefore a substitution
here as well, and not a cancellation between two divergent terms.

The angular equation is Fuchsian, and the substitution $z := \sin^{2}\theta$ makes that
visible. Reading $\Dth = \Xia + \lambda(b^{2}-a^{2})z$ and
$\chit = a^{2} + (b^{2}-a^{2})z$ as functions of $z$,
\begin{equation} \label{eq:KGfuchs}
  \frac{\dd^{2}\Theta}{\dd z^{2}}
  + \left( \frac{1}{z} + \frac{1}{z-1} + \frac{1}{z-d} \right)
    \frac{\dd\Theta}{\dd z}
  - \mathcal{G}(z)\,\Theta = 0 \,,
\end{equation}
with
\begin{equation} \label{eq:KGG}
  \mathcal{G}(z) := \frac{1}{4\Dth\,z(1-z)}
    \left[ \frac{\Xia m_1^{2}}{z} + \frac{\Xib m_2^{2}}{1-z}
         - \left( \eps + \frac{\Xia\Xib\,\Ecal^{2}}{\Dth} \right)\!\chit - K \right] ,
  \qquad
  d := -\frac{\Xia}{\lambda\,(b^{2}-a^{2})} \,.
\end{equation}
This is Eq.~(23) of \cite{Kunduri:2005fq} with the constant of \eqref{eq:KvsKKL} moved.
Theirs carries $1/\lambda$ in three separate places, a term $\mu^{2}/[4\lambda z(1-z)]$, the
$\Ecal^{2}$ term, and the $k' = k - \mu^{2}/\lambda$ hidden in their eigenvalue. The three
cancel only together. Equation \eqref{eq:KGG} carries none of them.

The singular points of \eqref{eq:KGfuchs} are $z = 0$, $z = 1$, $z = d$ and the point at
infinity. All four are regular, so the equation is of Heun type. Two limits send $d$ to
infinity, and they are not the same confluence.

The first is $a = b$. Then $\Dth \to \Xia$ and $\chit \to a^{2}$, both constant, the fourth
singular point is gone and the point at infinity stays regular. Three regular singular
points remain and the solutions are hypergeometric, which is the simplification recorded in
Eqs.~(24)--(26) of \cite{Kunduri:2005fq}. The indices are $\pm|m_1|/2$ at $z = 0$ and
$\pm|m_2|/2$ at $z = 1$, and at infinity they are $\gamma_1$ and $1-\gamma_1$ with
\begin{equation} \label{eq:KGindicial}
  \gamma_1\,(\gamma_1 - 1)
  = \frac{K + a^{2}\bigl( \eps + \Xia\,\Ecal^{2} \bigr)}{4\,\Xia} \,.
\end{equation}
This is Eq.~(25) of \cite{Kunduri:2005fq}, again with the poles at $\lambda = 0$ gone. It is
unchanged under $\gamma_1 \to 1-\gamma_1$, so it does not matter which of the two indices is
used. Requiring regularity at both ends truncates the hypergeometric series at degree $n$,
which fixes $\gamma_1$ and with it the eigenvalue,
\begin{equation} \label{eq:KGeigen}
  K = 4\,\Xia\,\gamma_1(\gamma_1 - 1) - a^{2}\bigl( \eps + \Xia\,\Ecal^{2} \bigr) \,,
  \qquad
  \gamma_1 = n + 1 + \frac{|m_1| + |m_2|}{2} \,,
  \qquad n = 0, 1, 2, \ldots
\end{equation}
Its Myers--Perry value is the substitution $\lambda = 0$, namely
$K = 4\gamma_1(\gamma_1-1) - a^{2}(\eps + \omega^{2})$.

The second limit is $\lambda \to 0$ at $a \neq b$, and it is of a different kind. Here
$\Dth \to 1$ while $\chit$ stays linear in $z$, so $z^{2}\mathcal{G}$ grows like
$(a^{2}-b^{2})(\mu^{2}-\omega^{2})z/4$ and the point at infinity becomes irregular. The
limit of \eqref{eq:KGfuchs} is a confluent Heun equation, the spheroidal equation with
spheroidicity $(a^{2}-b^{2})(\mu^{2}-\omega^{2})$, and not a hypergeometric one. What the
choice \eqref{eq:id12} buys is regularity of the coefficients at $\lambda = 0$, not of the
singularity structure. The Myers--Perry limit of the angular equation is a genuine
confluence, and no choice of separation constant can make it otherwise.

Returning to the radial equation, the consequence of the correction is not merely a change
of coefficients. A simple root $r_+$ of
$\Delta := V-2M$ is a regular singular point of \eqref{eq:KGrad}, near which the solutions
behave as $R \sim (r-r_+)^{\sigma}$: writing $y := r-r_+$ and keeping the leading terms,
\eqref{eq:KGrad} reduces to $y^{2}R'' + yR' - \bigl[A/\Delta'(r_+)\bigr]R = 0$ with
$A := \lim_{r \to r_+}(r-r_+)\widetilde{V}$, so that the indicial exponents are given by
$\sigma^{2} = A/\Delta'(r_+)$. Since $V(r_+) = 2M$,
\begin{equation} \label{eq:sigmasq}
  \sigma^{2} = -\,\frac{4M^{2}\,\Psi(r_+)^{2}}{\bigl[\Delta'(r_+)\bigr]^{2}} \,.
\end{equation}
The identity \eqref{eq:key} collapses $\Psi$ at the horizon: writing
$\Psi = \Xir^{-1}\bigl[\,\omega - \sum_{i=1}^{2} a_i \Xii m_i/\Ri\,\bigr]$ and recognizing
$\Omega_i = a_i \Xii/(r_+^{2}+a_i^{2})$ as the angular velocities of the horizon,
\begin{equation} \label{eq:horexp}
  \Psi(r_+) = \frac{\omega - \Omega_i m_i}{\Xir(r_+)} \,,
  \qquad \mbox{so} \qquad
  \sigma = \pm\,\frac{2iM\,\bigl(\omega - \Omega_i m_i\bigr)}
                    {\Xir(r_+)\,\Delta'(r_+)} \,.
\end{equation}
The corrected potential thus reproduces the superradiance combination
$\omega - \Omega_i m_i$ automatically, as it must. That of Eq.~(16) of
\cite{Kunduri:2005fq} does not: the discrepancy \eqref{eq:Vexcess} has a simple pole at
$r_+$ as well, and shifts $\sigma^{2}$ by an $O(1)$ amount.
The ingoing boundary
condition, and with it the superradiance criterion, therefore cannot be stated correctly
from Eqs.~(16) or (27) of \cite{Kunduri:2005fq}. The behavior as $r \to \infty$ is
unaffected, since \eqref{eq:Vexcess} decays fast enough there to leave the asymptotically
anti-de~Sitter falloff exponents unchanged.

%%%%%%%%%%%%%%%%%%%%%%%%%%%%%%%%%%%%%%%%%%%%%%%%%%%%%%%%%%%%%%%%%%%%%%%%%%%%%%
%%  BIBLIOGRAPHY
%%%%%%%%%%%%%%%%%%%%%%%%%%%%%%%%%%%%%%%%%%%%%%%%%%%%%%%%%%%%%%%%%%%%%%%%%%%%%%


\begin{thebibliography}{99}

%\cite{Hackmann:thesis}
\bibitem{Hackmann:thesis}
E.~Hackmann,
``Geodesic equations in black hole space-times with cosmological constant,''
Ph.D.\ thesis, Universit\"at Bremen (2010).

%\cite{Cieslik:2022uki}
\bibitem{Cieslik:2022uki}
A.~Cie\'slik and P.~Mach,
``Revisiting timelike and null geodesics in the Schwarzschild spacetime: general
expressions in terms of Weierstrass elliptic functions,''
\mbox{\href{https://dx.doi.org/10.1088/1361-6382/ac95f2}{{\em Class.\ Quant.\ Grav.} {\bfseries 39}, 225003 (2022)}};
\arxiv{2203.12401}.

%\cite{Cieslik:2023qdc}
\bibitem{Cieslik:2023qdc}
A.~Cie\'slik, E.~Hackmann and P.~Mach,
``Kerr Geodesics in Terms of Weierstrass Elliptic Functions,''
\mbox{\href{https://dx.doi.org/10.1103/PhysRevD.108.024056}{{\em Phys.\ Rev.\ D} {\bfseries 108}, 024056 (2023)}};
\arxiv{2305.07771}.

%\cite{Biermann:1865}
\bibitem{Biermann:1865}
G.~G.~A.~Biermann,
``Problemata quaedam mechanica functionum ellipticarum ope soluta,''
Ph.D.\ thesis, Friedrich-Wilhelms-Universit\"at zu Berlin (1865).

%\cite{WhittakerWatson}
\bibitem{WhittakerWatson}
E.~T.~Whittaker and G.~N.~Watson,
{\em A Course of Modern Analysis}, 4th ed.
(Cambridge University Press, Cambridge, 1927).

%\cite{Greenhill:1892}
\bibitem{Greenhill:1892}
A.~G.~Greenhill,
{\em The Applications of Elliptic Functions}
(Macmillan and Co., London, 1892).

%\cite{Frolov:2017kze}
\bibitem{Frolov:2017kze}
V.~P.~Frolov, P.~Krtou\v{s} and D.~Kubiz\v{n}\'ak,
``Black holes, hidden symmetries, and complete integrability,''
\mbox{\href{https://dx.doi.org/10.1007/s41114-017-0009-9}{{\em Living Rev.\ Rel.} {\bfseries 20}, 6 (2017)}};
\arxiv{1705.05482}.

%\cite{Kunduri:2005fq}
\bibitem{Kunduri:2005fq}
H.~K.~Kunduri and J.~Lucietti,
``Integrability and the Kerr-(A)dS black hole in five dimensions,''
\mbox{\href{https://dx.doi.org/10.1103/PhysRevD.71.104021}{{\em Phys.\ Rev.\ D} {\bfseries 71}, 104021 (2005)}};
\arxiv{hep-th/0502124}.

%\cite{Grunau:2017uwv}
\bibitem{Grunau:2017uwv}
S.~Grunau, H.~Neumann and S.~Reimers,
``Geodesic motion in the five-dimensional Myers-Perry-AdS spacetime,''
\mbox{\href{https://dx.doi.org/10.1103/PhysRevD.97.044011}{{\em Phys.\ Rev.\ D} {\bfseries 97}, 044011 (2018)}};
\arxiv{1711.02933}.

%\cite{Hawking:1998kw}
\bibitem{Hawking:1998kw}
S.~W.~Hawking, C.~J.~Hunter and M.~Taylor,
``Rotation and the AdS / CFT correspondence,''
\mbox{\href{https://dx.doi.org/10.1103/PhysRevD.59.064005}{{\em Phys.\ Rev.\ D} {\bfseries 59}, 064005 (1999)}};
\arxiv{hep-th/9811056}.

%\cite{Gibbons:2004uw}
\bibitem{Gibbons:2004uw}
G.~W.~Gibbons, H.~L\"u, D.~N.~Page and C.~N.~Pope,
``The General Kerr-de Sitter metrics in all dimensions,''
\mbox{\href{https://dx.doi.org/10.1016/j.geomphys.2004.05.001}{{\em J.\ Geom.\ Phys.} {\bfseries 53}, 49--73 (2005)}};
\arxiv{hep-th/0404008}.

%\cite{Davis:2005ys}
\bibitem{Davis:2005ys}
P.~Davis, H.~K.~Kunduri and J.~Lucietti,
``Special symmetries of the charged Kerr-AdS black hole of D=5 minimal gauged supergravity,''
\mbox{\href{https://dx.doi.org/10.1016/j.physletb.2005.09.062}{{\em Phys.\ Lett.\ B} {\bfseries 628}, 275--280 (2005)}};
\arxiv{hep-th/0508169}.

%\cite{Wu:2008bc}
\bibitem{Wu:2008bc}
S.~Q.~Wu,
``Symmetry operators and separability of the massive Dirac's equation in the general
5-dimensional Kerr-(anti-)de Sitter black hole background,''
\mbox{\href{https://dx.doi.org/10.1088/0264-9381/26/5/055001}{{\em Class.\ Quant.\ Grav.} {\bfseries 26}, 055001 (2009)}};
[erratum: \mbox{\href{https://dx.doi.org/10.1088/0264-9381/26/18/189801}{{\em Class.\ Quant.\ Grav.} {\bfseries 26}, 189801 (2009)}}];
\arxiv{0808.3435}.

%\cite{Vasudevan:2005js}
\bibitem{Vasudevan:2005js}
M.~Vasudevan and K.~A.~Stevens,
``Integrability of particle motion and scalar field propagation in Kerr-(Anti) de Sitter
black hole spacetimes in all dimensions,''
\mbox{\href{https://dx.doi.org/10.1103/PhysRevD.72.124008}{{\em Phys.\ Rev.\ D} {\bfseries 72}, 124008 (2005)}};
\arxiv{gr-qc/0507096}.

%\cite{Li:2009zzs}
\bibitem{Li:2009zzs}
H.~L.~Li, W.~Y.~Qi and R.~Lin,
``Hawking radiation of Dirac particles from Kerr-AdS black hole in five dimensions,''
\mbox{\href{https://dx.doi.org/10.1016/j.physletb.2009.05.044}{{\em Phys.\ Lett.\ B} {\bfseries 677}, 332--337 (2009)}}.

%\cite{Frolov:2003en}
\bibitem{Frolov:2003en}
V.~P.~Frolov and D.~Stojkovic,
``Particle and light motion in a space-time of a five-dimensional rotating black hole,''
\mbox{\href{https://dx.doi.org/10.1103/PhysRevD.68.064011}{{\em Phys.\ Rev.\ D} {\bfseries 68}, 064011 (2003)}};
\arxiv{gr-qc/0301016}.

%\cite{Mino:2003yg}
\bibitem{Mino:2003yg}
Y.~Mino,
``Perturbative approach to an orbital evolution around a supermassive black hole,''
\mbox{\href{https://dx.doi.org/10.1103/PhysRevD.67.084027}{{\em Phys.\ Rev.\ D} {\bfseries 67}, 084027 (2003)}};
\arxiv{gr-qc/0302075}.

%\cite{Hackmann:2010zz}
\bibitem{Hackmann:2010zz}
E.~Hackmann, C.~L\"ammerzahl, V.~Kagramanova and J.~Kunz,
``Analytical solution of the geodesic equation in Kerr-(anti) de Sitter space-times,''
\mbox{\href{https://dx.doi.org/10.1103/PhysRevD.81.044020}{{\em Phys.\ Rev.\ D} {\bfseries 81}, 044020 (2010)}};
\arxiv{1009.6117}.

%\cite{Diemer:2014lba}
\bibitem{Diemer:2014lba}
V.~Diemer, J.~Kunz, C.~L\"ammerzahl and S.~Reimers,
``Dynamics of test particles in the general five-dimensional Myers-Perry spacetime,''
\mbox{\href{https://dx.doi.org/10.1103/PhysRevD.89.124026}{{\em Phys.\ Rev.\ D} {\bfseries 89}, 124026 (2014)}};
\arxiv{1404.3865}.

%\cite{Faraji:2025xkl}
\bibitem{Faraji:2025xkl}
S.~Faraji and E.~Hackmann,
``Null integrability and photon phenomenology in the accelerated Schwarzschild black hole,''
\mbox{\href{https://dx.doi.org/10.1103/ffnb-n4nj}{{\em Phys.\ Rev.\ D} {\bfseries 113}, 124030 (2026)}};
\arxiv{2509.23432}.

%\cite{Page:2006ka}
\bibitem{Page:2006ka}
D.~N.~Page, D.~Kubiz\v{n}\'ak, M.~Vasudevan and P.~Krtou\v{s},
``Complete integrability of geodesic motion in general Kerr-NUT-AdS spacetimes,''
\mbox{\href{https://dx.doi.org/10.1103/PhysRevLett.98.061102}{{\em Phys.\ Rev.\ Lett.} {\bfseries 98}, 061102 (2007)}};
\arxiv{hep-th/0611083}.

%\cite{Frolov:2006pe}
\bibitem{Frolov:2006pe}
V.~P.~Frolov, P.~Krtou\v{s} and D.~Kubiz\v{n}\'ak,
``Separability of Hamilton-Jacobi and Klein-Gordon Equations in General Kerr-NUT-AdS
Spacetimes,''
\mbox{\href{https://dx.doi.org/10.1088/1126-6708/2007/02/005}{{\em JHEP} {\bfseries 02}, 005 (2007)}};
\arxiv{hep-th/0611245}.

%\cite{Chong:2005hr}
\bibitem{Chong:2005hr}
Z.~W.~Chong, M.~Cveti\v{c}, H.~L\"u and C.~N.~Pope,
``General non-extremal rotating black holes in minimal five-dimensional gauged
supergravity,''
\mbox{\href{https://dx.doi.org/10.1103/PhysRevLett.95.161301}{{\em Phys.\ Rev.\ Lett.} {\bfseries 95}, 161301 (2005)}};
\arxiv{hep-th/0506029}.

%\cite{Drawer:2020mpw}
\bibitem{Drawer:2020mpw}
J.-C.~Drawer and S.~Grunau,
``Geodesic motion around a supersymmetric AdS$_5$ black hole,''
\mbox{\href{https://dx.doi.org/10.1140/epjc/s10052-020-8101-9}{{\em Eur.\ Phys.\ J.\ C} {\bfseries 80}, 536 (2020)}};
\arxiv{2002.05481}.

\end{thebibliography}
\end{document}